\documentclass[preprint]{aastex631}

\usepackage{multirow}
\usepackage{bm}
\usepackage{amsmath}

\shorttitle{Photometric [Fe/H] of RRab stars in the $G$ and $K_s$ bands by deep learning}
\shortauthors{D\'ek\'any et al.}
\graphicspath{{./}{figures/}}

\begin{document}

\title{Photometric metallicity prediction of fundamental-mode RR~Lyrae stars in the Gaia optical and $K_s$ infrared wavebands by deep learning
}

\author[0000-0001-7696-8331]{Istv\'an D\'ek\'any}
\affiliation{Astronomisches Rechen-Institut, 
	Zentrum f\"ur Astronomie der Universit\"at Heidelberg,
	M\"onchhofstr. 12-14, 69120 Heidelberg, Germany}

\author[0000-0002-1891-3794]{Eva K. Grebel}
\affiliation{Astronomisches Rechen-Institut, 
	Zentrum f\"ur Astronomie der Universit\"at Heidelberg,
	M\"onchhofstr. 12-14, 69120 Heidelberg, Germany}



\begin{abstract}

RR~Lyrae stars are useful chemical tracers thanks to the empirical relationship between their heavy-element abundance and the shape of their light curves. However, the consistent and accurate calibration of this relation across multiple photometric wavebands has been lacking.
We have devised a new method for the metallicity estimation of fundamental-mode RR~Lyrae stars in the Gaia optical $G$ and near-infrared VISTA $K_s$ wavebands by deep learning. First, an existing metallicity prediction method is applied to large photometric data sets, which are then used to train long short-term memory recurrent neural networks for the regression of the [Fe/H] to the light curves in other wavebands. This approach allows an unbiased transfer of our accurate, spectroscopically calibrated $I$-band formula to additional bands at the expense of minimal additional noise. We achieve a low mean absolute error of $0.1$~dex and high $R^2$ regression performance of $0.84$ and $0.93$ for the $K_s$ and $G$ bands, respectively, measured by cross-validation. The resulting predictive models are deployed on the Gaia DR2 and VVV inner-bulge RR~Lyrae catalogs. We estimate mean metallicities of $-1.35$~dex for the inner bulge and $-1.7$ for the halo, which are significantly less than values obtained by earlier photometric prediction methods. Using our results, we establish a public catalog of photometric metallicities of over 60,000 Galactic RR~Lyrae stars, and provide an all-sky map of the resulting RR~Lyrae metallicity distribution. The software code used for training and deploying our recurrent neural networks is made publicly available in the open-source domain.

\end{abstract}



\section{Introduction} \label{sec:intro}

RR~Lyrae variables are low-mass, core helium-burning stars undergoing radial pulsation. Since these bright stars abound in old stellar populations, are easy to identify, and their absolute luminosities and heavy-element abundances can be inferred from their light variations \citep[see][for a recent review]{2022arXiv220206982B}, they have been widely employed as astrophysical tracer objects within the Local Group of galaxies \citep[e.g.,][]{2018MNRAS.478.4590T,clementini_gaia_2019,soszynski_over_2019,dekany_near-infrared_2020}. In addition, the photometric estimation of the metallicity distributions of large RR~Lyrae samples offer the possibility to constrain the formation epoch and early chemical enrichment of the underlying old stellar populations \citep[see, e.g.,][and references therein]{savino_age_2020}.

Although the existence of an empirical relationship between the shape of an RR~Lyrae star's light curve and its metallicity has been known since the pioneering work by \citet{jurcsik_determination_1996}, its accurate calibration in multiple photometric bands has been a challenging task, mainly due to the small number of stars with high-dispersion spectroscopic (HDS) metallicity measurements. Over the past $\sim25$ years, various empirical formulae have been established to predict the [Fe/H] from the Fourier regression parameters of the light curves in various photometric wavebands, following different strategies. Since direct regression to the HDS metallicities has long suffered from small sample sizes \citep[e.g.,][]{nemec_metal_2013}, most authors \citep[e.g.,][]{jurcsik_determination_1996,smolec_metallicity_2005,ngeow_palomar_2016,iorio_chemo-kinematics_2020,mullen_metallicity_2021} addressed the problem by relying on low-dispersion spectroscopic or spectro-photometric metallicities estimated from spectral indices \citep[see, e.g.,][]{layden_metallicities_1994,crestani_deltaS}. 
Although the diversity of light-curve shapes in the training sets could be increased by a few times this way, such an approach required the intermediate calibration of the spectral indices to the same limited amount of HDS data. Another approach was to transfer a predictive formula established for one waveband to another by linearly transforming the regressors, i.e., the Fourier parameters of the light curve, between these bands \citep[e.g.,][]{skowron_ogle-ing_2016,clementini_gaia_2019}. However, such transformations are not only inherently noisy, but they also depend on the metallicity.

Generally speaking, the limited amount of heterogeneous data, strong heteroskedasticity, and large errors in the regressors lead to various systematic biases in the metallicity prediction formulae. Such systematics were often also difficult to explore and quantify due to the lack of large samples with accurate light curves in multiple wavebands. Fortunately, HDS analyses of RR~Lyrae stars have recently proliferated \citep[see][and references therein]{crestani_deltaS}, calling for the revision of earlier photometric [Fe/H] estimation methods. Exploiting the new state-of-the-art abundance measurements, we recently established new empirical predictive models of the [Fe/H] of both fundamental-mode (RRab) and first-overtone (RRc) RR~Lyrae stars from their light curves in the Cousins $I$ filter, a highly important waveband in contemporary studies of the Milky Way and the Magellanic Clouds. In the case of this band, the training sample was sufficiently large and accurate, so that with careful feature selection and probabilistic modeling of the uncertainties, it allowed a direct regression to the HDS data, removing the additional error-prone but customary intermediate step of calibrating spectral indices to the HDS [Fe/H] values. Our analysis revealed large systematic biases in several earlier formulae, that led to an $\sim0.4$\,dex positive bias in previous estimates of the metallicity distribution functions of old stellar populations, such as the bulge and the Magellanic Clouds.

In the era of large time-domain sky surveys that discovered and monitor $\sim10^5$ RR~Lyrae stars in the Local Group and keep on detecting new ones, accurate photometric predictive models of their metallicties are needed in order to unlock their full potential as population tracers; and ideally we need them in all the various optical and infrared wavebands in which such surveys operate. One way to accomplish that is to consider this task as a set of independent regression problems of the HDS measurements on the light curves in each waveband. However, this approach is hindered by the fact that in most cases, the number of stars with both HDS abundance measurements and accurate light curves is still too small. Using small, separate training data sets for different wavebands is not only prone to systematics in the predicted [Fe/H], but also makes it hard to keep all separate relations up-to-date as new data become available.

An alternative approach is to fit a single predictive formula to the HDS measurements using a waveband in which a sufficiently large HDS training data set already exists. Then, transfer the resulting {\em base estimator} to other wavebands; but instead of transforming the regressors, handle each problem as a regression of the light curve shapes to the predicted {\em photometric} [Fe/H] as the response variable. By taking this approach, large training data sets become available, allowing us to use deep learning for the regression, i.e., highly complex models capable of extracting every last bit of predictive information from the light curves. This, in turn, offers the possibility of a highly precise and unbiased transfer of the original relation to additional wavebands, with the cost of introducing only minimal noise. Therefore, the safe combination of photometric data across different surveys for the inference of metallicity distribution functions of large stellar populations becomes possible. Moreover, upon the availability of additional HDS measurements, an immediate homogeneous update of the predictive models in all wavebands becomes very straightforward.

It is important to emphasize that by adopting this method, we make the important assumption of physical similarity between the HDS training data set of the base estimator and the training sets of the deep learned models. Specifically, we assume that these data sets are not systematically different in terms of other, latent physical variables that affect the light-curve shapes (e.g., in their helium abundances; see, e.g., \citealt{2021ApJ...920...33D}). We note that the same assumption is also implicitly made when any other estimator of photometric metallicity estimation is either calibrated or applied to a new data set.

The above approach of photometric metallicity estimation was originally put forward by \citet{hajdu_data-driven_2018}, who trained an ensemble of multi-layer perceptrons (i.e., classical fully connected neural networks) to predict the [Fe/H] of RRab stars from their $K_s$-band photometric time series, using a parametric representation of the light curves as regressors, and $I$-band photometric [Fe/H] estimates obtained with the formula of \citet{smolec_metallicity_2005} as response variables. In this study, we build upon this idea to obtain deep-learned predictive models of the [Fe/H] of RRab stars from their light curves in the Gaia $G$ and VISTA $K_s$ wavebands. Our analysis comprises the following major steps. First, using our recent predictive formula \citep[][hereafter D21]{2021ApJ...920...33D}, we create our development data sets by computing photometric [Fe/H] estimates from the $I$-band light curves of a large number of RRab stars that were also observed in either the Gaia optical $G$ or the VISTA near-infrared (near-IR) $K_s$ filters. These [Fe/H]$_I$ estimates are then used as our response variables for the development of additional predictive models. To estimate the [Fe/H]$_I$ from $G$ and $K_s$ light curves, we train Recurrent Neural Networks (RNNs), which allow us to use the original photometric time series as input data, instead of relying on a classical parametric representation of the light curves. Finally, the trained RNNs are deployed on the RRab catalogs of the Gaia and the VISTA Variables in the V\'ia L\'actea surveys, in order to deliver a public database of accurate photometric metallicity estimates.

\section{Photometric data and their representation}\label{sec:data}

As the first step of establishing the photometric data sets for the development of our predictive models for the metallicity, we compute $I$-band photometric [Fe/H] estimates for a very large number of Galactic RRab stars using the D21 formula, that is directly calibrated to HDS spectroscopic measurements. These [Fe/H]$_I$ estimates will serve as our response variables for training and validating the predictive models for the optical and near-IR bands. We used the public $I$-band light curves of RRab stars acquired by the Optical Gravitational Lensing Experiment \citep[OGLE,][]{udalski_ogle-iv_2015} cataloged in the OGLE Collection of Variable Stars\footnote{\url{http://ogledb.astrouw.edu.pl/~ogle/OCVS/}} (OCVS) of the Galactic bulge and disk, comprising tens of thousands of objects.

The pulsation periods of the RRab stars were adopted from the OCVS, while the $I$-band $\phi_{31}$ and $A_2$ Fourier parameters that also serve as input features of the D21 formula were computed using the \texttt{lcfit}\footnote{\url{https://zenodo.org/record/6576222}} package \citep{lcfit}, following the regression procedure discussed by \citet{2021ApJ...920...33D}. In brief, a robust fitting of a truncated Fourier-sum with iterative outlier rejection and order optimization via cross-validation is followed by a Gaussian Process Regression (GPR) of the phase-folded light curves. The final parameters are obtained from the Fourier representation of the mean GPR model. The results of our analysis are displayed in Table~\ref{tab:ifeh_ocvs}. The uncertainties in the resulting [Fe/H]$_I$ values were estimated by drawing numerous random samples from the GPR model at the original observational phases for each object, repeating the same regression procedure as above for each realization, and calculating the standard deviation of the resulting [Fe/H] predictions. These uncertainties will be used for sample weighting in the training and validation of the predictive models.

\begin{splitdeluxetable*}{cccccBccccccccccD}
	\tablecaption{$I$-band photometric parameters and metallicities of RRab stars in the bulge and disk areas of the OCVS. \label{tab:ifeh_ocvs}}
	\tablehead{
		\colhead{Field\tablenotemark{a}} & \colhead{ID} & \colhead{R.A.\tablenotemark{b} [hms]} & \colhead{DEC.\tablenotemark{b} [dms]} & \colhead{[Fe/H]$_I$} & \colhead{$\langle I \rangle$} & \colhead{$N_{ep.}$} & 
		\colhead{Period~[d]} & \colhead{$A_{tot.}$} & \colhead{$A_{1}$} & \colhead{$A_2$} & \colhead{$A_3$} & \colhead{$\phi_{21}$} &
		\colhead{$\phi_{31}$} & \colhead{$C_{\varphi}$\tablenotemark{c}} & \multicolumn2c{$S/N$}
	}
	\startdata
	b & 00001 &  17:05:07.49  & -32:37:57.2  & -1.712 &  15.595  &  121 &  0.732586 &  0.447 &   0.167 &  0.073  & 0.043 &  9.5708 &  6.680 &  0.956 &  413.2 \\
	b & 00003 &  17:05:09.88  & -32:39:52.8  & -1.296 &  16.451  &  121 &  0.515465 &  0.696 &   0.215 &  0.118  & 0.077 &  8.9977 &  5.700 &  0.950 &  428.2 \\
	b & 00004 &  17:05:15.22  & -32:50:13.2  & -1.868 &  15.965  &  122 &  0.651696 &  0.264 &   0.113 &  0.036  & 0.018 &  9.3813 &  6.315 &  0.965 &  154.5 \\
	b & 00005 &  17:05:21.98  & -32:39:42.5  & -2.596 &  15.336  &  121 &  0.565511 &  0.606 &   0.241 &  0.098  & 0.029 &  8.9175 &  5.006 &  0.962 &  72.3  \\
	b & 00006 &  17:05:29.00  & -32:33:37.4  & -1.426 &  15.846  &  121 &  0.478877 &  0.804 &   0.245 &  0.121  & 0.097 &  8.8742 &  5.396 &  0.958 &  547.3 \\
	\enddata
	\tablenotetext{a}{`b': bulge, `d': disk; original OCVS designations: OGLE-BLG-RRLYR-ID and OGLE-GD-RRLYR-ID, respectively.}
	\tablenotetext{b}{Coordinates of J2000.0 epoch.}
	\tablenotetext{c}{Phase coverage: $1-$maximum phase lag.}
	\tablecomments{
		Results are shown for stars that pass the following criteria: $C_\varphi\geq0.8$, $A_{tot.}\leq1.2$, $S/N\geq50$.\newline
		This table is available in its entirety in machine-readable form.}
\end{splitdeluxetable*}

At the next step, we cross-matched the celestial coordinates of the OCVS RRab stars with the RR~Lyrae catalog from the Gaia data release 2 (DR2) \citep{clementini_gaia_2019}, and the $K_s$-band point source catalogs of the VISTA Variables in the V\'ia L\'actea (VVV) ESO Public Survey \citep{minniti_vista_2010} created by the VISTA Data Flow System \citep[VDFS,][]{2004SPIE.5493..401E} and provided by the Cambridge Astronomy Survey Unit (CASU). The cross-match yielded $\sim13,700$ objects with Gaia and $\sim29,600$ stars with VVV photometry.

The Gaia light curves of the cross-matched objects were directly retrieved through the Gaia@AIP database interface of the Leibniz Institute for Astrophysics Potsdam. The corresponding $K_s$-band VVV light curves were obtained by following the procedure discussed in \citet{dekany_near-infrared_2018}, and subsequently applying the photometric zero-point corrections of \citet{2020ExA....49..217H}.

Both the Gaia and VVV light curves were pre-processed by subjecting them to the previously discussed regression procedure, which was also used for the OGLE light curves. Since the sampling, precision, and temporal baseline of the OGLE light curves are far superior than those of the Gaia and VVV data, we always used the periods derived from the former. Likewise, we adopted the OGLE variable star classifications for similar reasons. In case of the Gaia data, we opted not to use the per-epoch photometric quality flags provided in DR2 for the omission of bad data, instead we relied on our own outlier rejection mechanism because the former would have led to the culling of too many useful measurements. In case of the VVV data, we selected the optimal photometric aperture for each object by maximizing the signal-to-noise ($S/N$) ratio in the cleaned light curve.

For the predictive modeling of the [Fe/H] from the light curves, we use the following 2-dimensional sequences as input variables (i.e., regressors):

\begin{eqnarray}\label{eq:input_seq}
\mathbf{x}^{<t>}&=& \left(\begin{array}{cc}{m^{<t>}-\langle m \rangle} \\ {\phi_P^{<t>} \cdot P}\end{array} \right),~t=\{1,\dots,N_{ep.}\}~,\\
\phi_P(T)&=&{\rm mod}[(T+P\cdot\Phi_1/(2\pi))/P]~,
\end{eqnarray}

\noindent Here, $m^{<t>}$ and $\phi_P^{<t>}$ are the magnitudes and corresponding pulsation phases of the light curve, respectively, $\langle m \rangle$ is the mean magnitude, $\Phi_1$ is the phase of the first Fourier term, $T$ is the observation time, and $N_{ep.}$ is the number of observational epochs. In addition, the VVV $K_s$ light curves were binned to a maximum of 60 points per time series in order to speed up the training of the neural networks.

By using the original (phased) measurements instead of a traditional parametric representation of the light curves, we virtually eliminate any potential representation error in the regressors, and allow a natural propagation of uncertainties in the photometric time series in the form of scatter in the input sequences. We note that the above representation of the time series is not strictly non-parametric due to the phase-folding of the light curves: an accurate knowledge of the period is obviously required or else the input sequence will suffer from bias. In addition, the input sequences are phase-aligned by $\Phi_1$, which thereby acts as a nuisance parameter, and its uncertainty can still lead to some representation error due to sequence misalignment. In practice, however, the robustness of this phase alignment method ensures negligible bias in the input for all but the noisiest and most ill-sampled light curves.

\begin{figure}
	\gridline{
		\fig{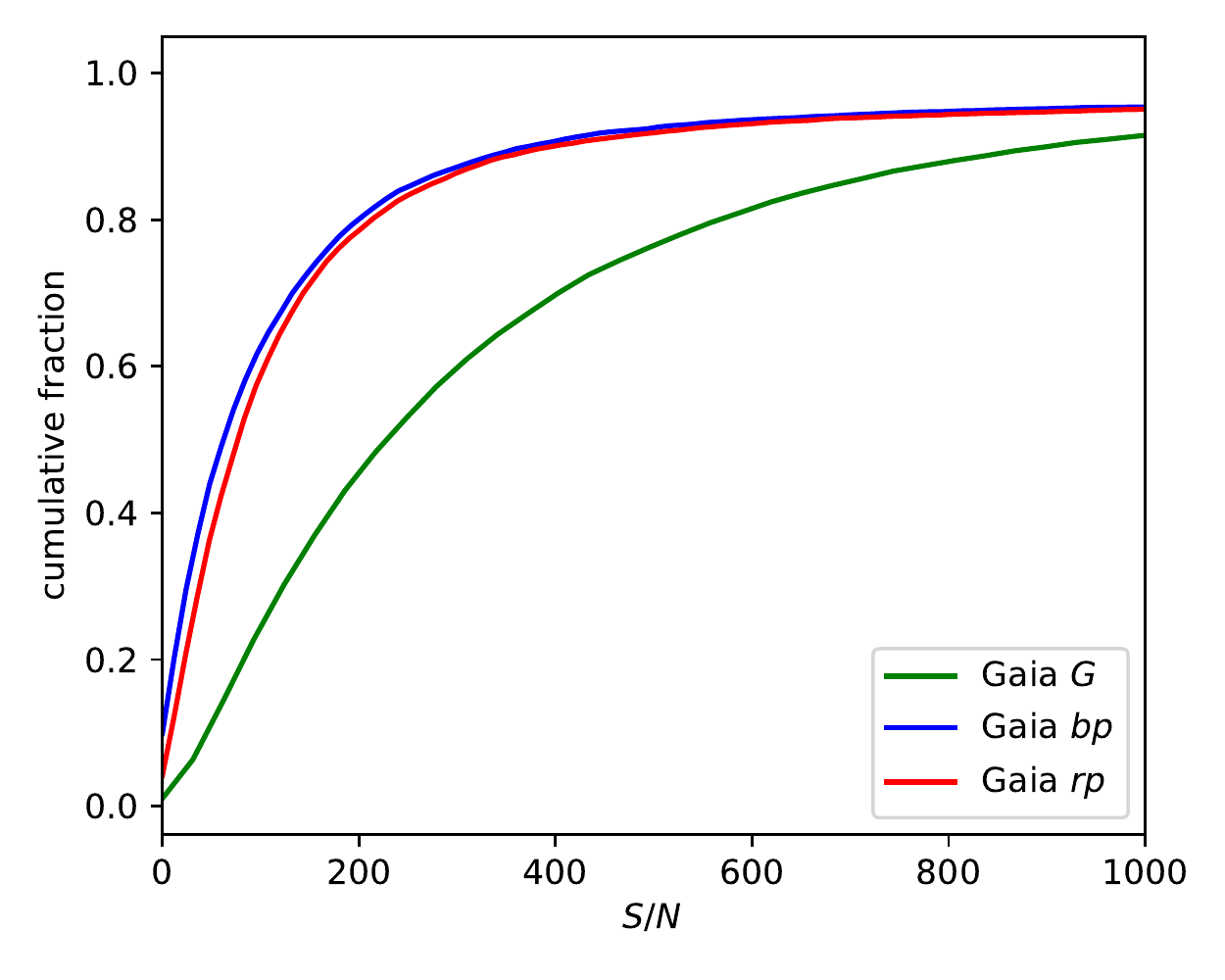}{0.4\textwidth}{}
	}
	\caption{Cumulative distributions of the $S/N$ of the Gaia light curves in the $G$, $bp$, and $rp$ wavebands for the Galactic RRab stars in the common sample with the OCVS.
		\label{fig:snr_cumul_hist}}
\end{figure}

In addition to the broad optical $G$ waveband, Gaia also acquires simultaneous photometric measurements in its $bp$ (blue) and $rp$ (red) filters. In principle, we could use all three time-series as 6-dimensional sequences (3 magnitudes and their corresponding phases) at the input of the RNN. However, the $bp$ and $rp$ light curves in the common OCVS--Gaia data set have much lower $S/N$ and significantly less useful data points than their $G$-band counterparts (Fig.~\ref{fig:snr_cumul_hist}). Therefore, they would mostly add noise to our model, and introducing quality criteria to counter this would shrink the data set too much. Moreover, bad data points in the different bands occur at different phases, thus their independent omission would disrupt the matching phase values across the different bands for the same star, the handling of which would require increased complexity of the RNN model. In view of the above points, we decided to rely solely on the $G$-band Gaia light curves in our analysis.

The obtained database of phase-folded, phase-aligned, and cleaned photometric time series was narrowed down by imposing various criteria for controlling their quality. Light curves with very few measurements, low $S/N$, poor phase coverage, and uncertain [Fe/H]$_I$ estimates were excluded from the regressions. We also trimmed outliers beyond the main locus of the period--amplitude--metallicity distributions. Formally, such criteria can be considered as tunable hyper-parameters whose optimal values maximize the predictive model's performance at validation (see Sect.~\ref{subsec:optimization}), and provide a good trade-off between data quality and quantity. Our final criteria for the development data sets are summarized below:\\

{\bf Gaia} $G$:
\begin{equation}\label{eq:crit_G}
 C_{\varphi}>0.85;
~S/N>30;
~N_{ep.}>20;~A_{tot.}<1.4;~-2.7<[{\rm Fe}/{\rm H}]_I<0;
~\sigma_{[{\rm Fe}/{\rm H}]_I}<0.3 
\end{equation}

{\bf VVV} $K_s$:
\begin{equation}\label{eq:crit_Ks}
C_{\varphi}>0.9;~S/N>100;
~N_{ ep.}>100;~A_{tot.}<0.45;~-2.5<[{\rm Fe}/{\rm H}]_I;
~\sigma_{[{\rm Fe}/{\rm H}]_I}<0.15;
~P_{Bl.}<0.7
\end{equation}

\noindent Here, $C_{\varphi}$ denotes the phase coverage (i.e., $1-$maximum phase lag), $A_{tot.}$ is the total (peak-to-valley) amplitude of the GPR light-curve model, $N_{ep.}$ is the number of epochs in the light curve, and $\sigma_{[{\rm Fe}/{\rm H}]_I}$ is the uncertainty in the [Fe/H]$_I$ values, estimated from the GPR model of the $I$-band light curve. Moreover, $P_{Bl.}$ is the probability that a star shows the Blazhko effect \citep[see, e.g., ][for a review]{2016pas..conf...22S}, i.e., a periodic amplitude and/or phase modulation of the light curve according to the machine-learned classifier of \citet{2017MNRAS.466.2602P}.
Our selection criterion excludes stars with strongly modulated light curves from the $K_s$ development set. 
Strong phase modulation due to the Blazhko effect, especially when undersampled, biases the RRab stars' light-curve shape and thus affects their photometric metallicity prediction \citep[see, e.g.,][]{nemec_metal_2013,2021ApJ...920...33D}. Even though the strength of the modulation is generally larger in the optical than in the near-IR \citet{2018MNRAS.475.4208J} for the same star, introducing a selection criterion by $P_{Bl.}$ for the $G$-band data, unlike for the $K_s$ band, did not give any advantage. The much more subtle dependence of the light curve shape on the metallicity in the $K_s$ band and the highly non-uniform sampling of the VVV light curves probably both contribute to this behavior. Finally, we note that we imposed limits on the input metallicity range in order to exclude possibly spurious [Fe/H]$_I$ values. The applied limits trim the lowest 1 percentiles of the distributions and remove only 7 objects with positive $I$-band metallicities from the $G$-band data set.

\begin{figure*}
	\gridline{
		\fig{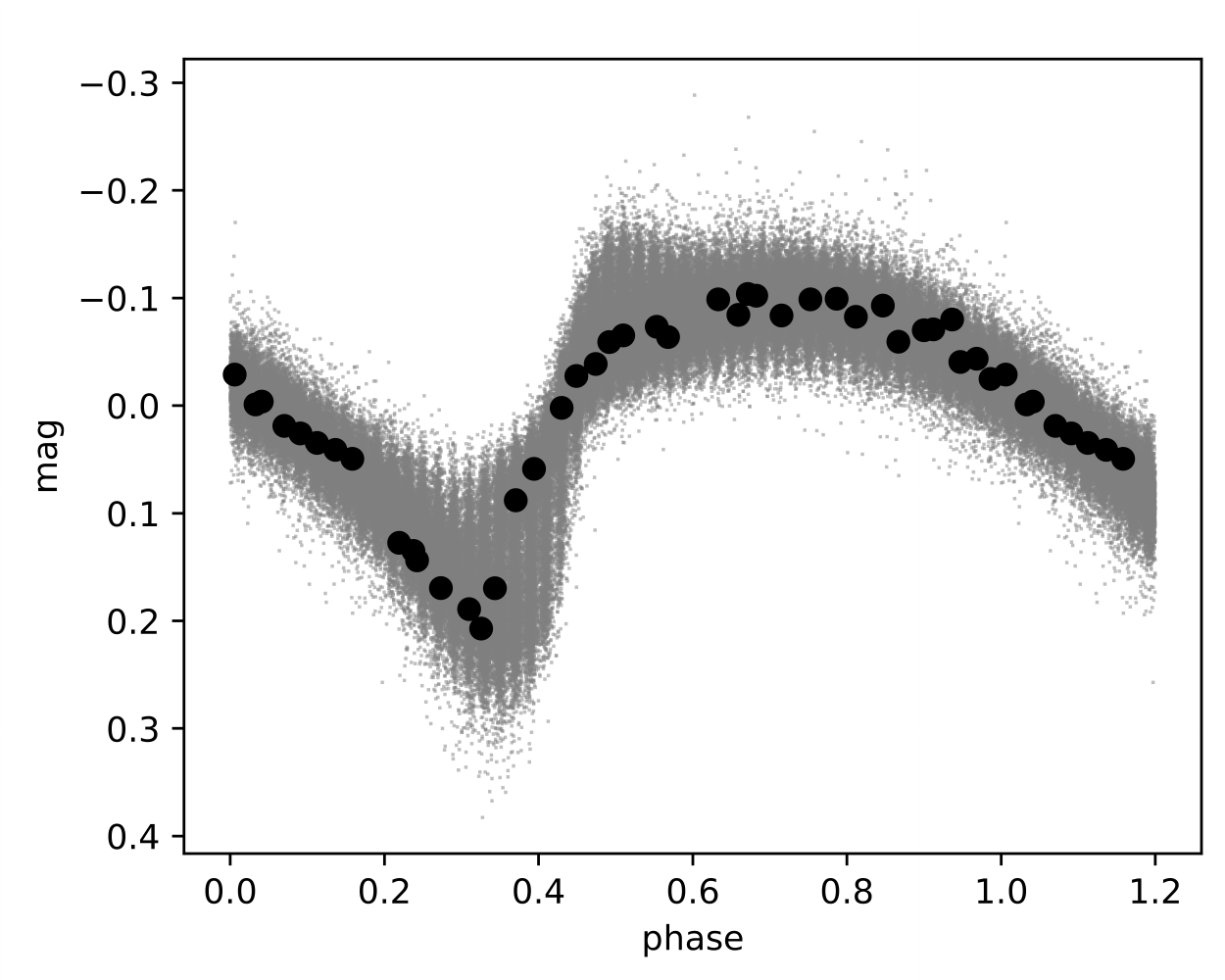}{0.45\textwidth}{}
		\hskip-0.8cm
		\fig{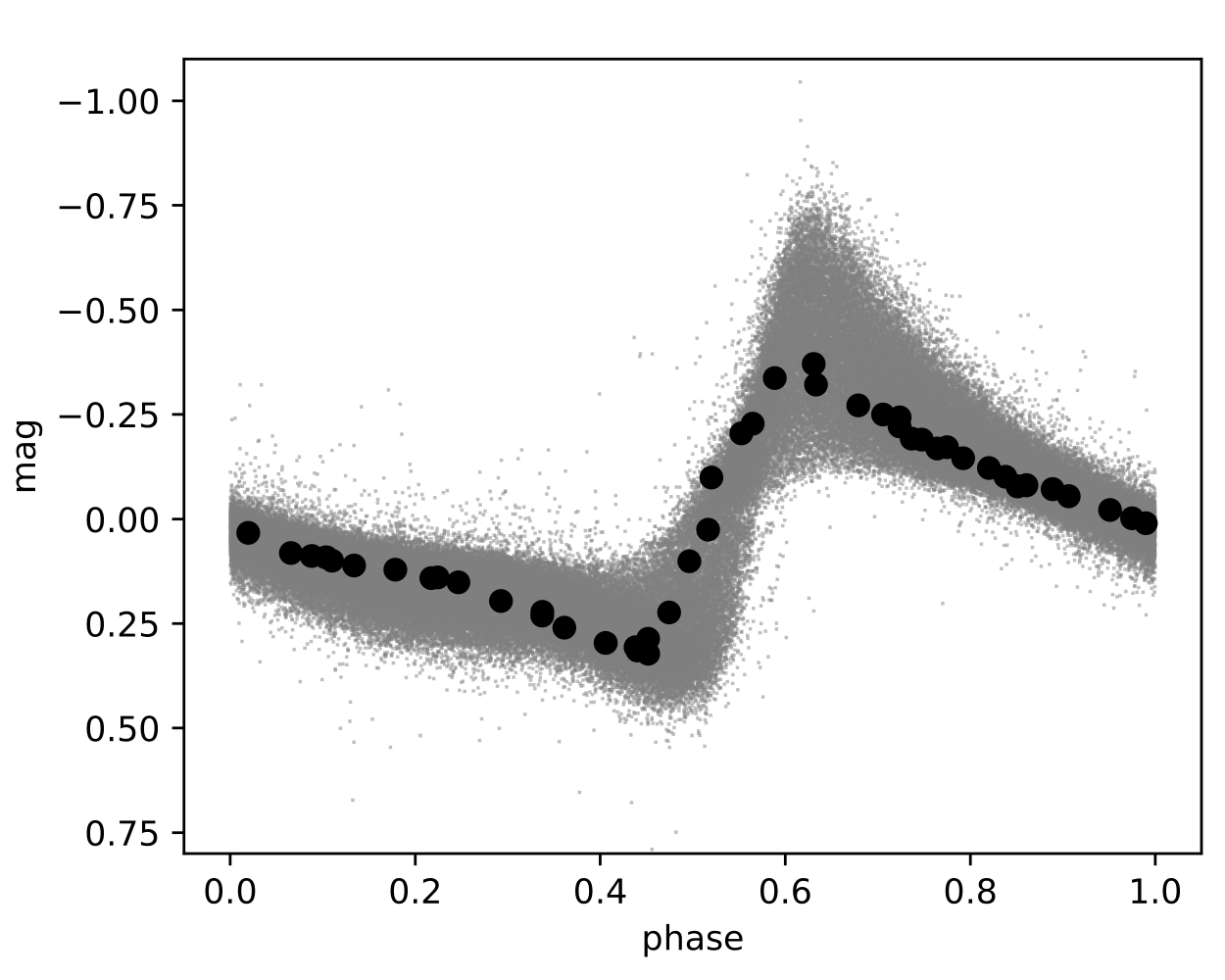}{0.45\textwidth}{}
	}
	\caption{Phase-folded, phase-aligned, mean-subtracted light curves of the $K_s$-band (left) and $G$-band (right) development data sets after all pre-processing steps have been applied. Gray dots show all data, black points highlight the data for a single RRab star. We note that the vertical structures in the left panel are due to data binning (see text).
		\label{fig:lightcurves}}
\end{figure*}

\begin{figure*}
	\gridline{
		\fig{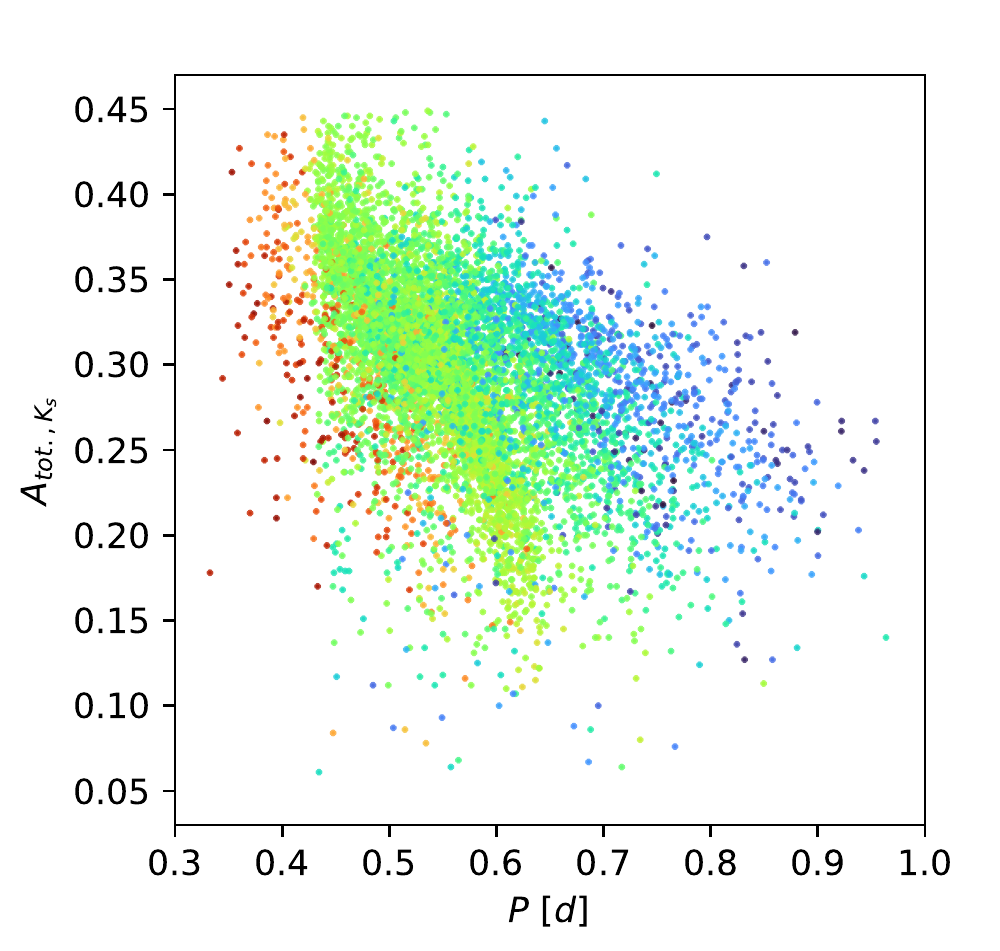}{0.4\textwidth}{}
		\hskip-0.8cm
		\fig{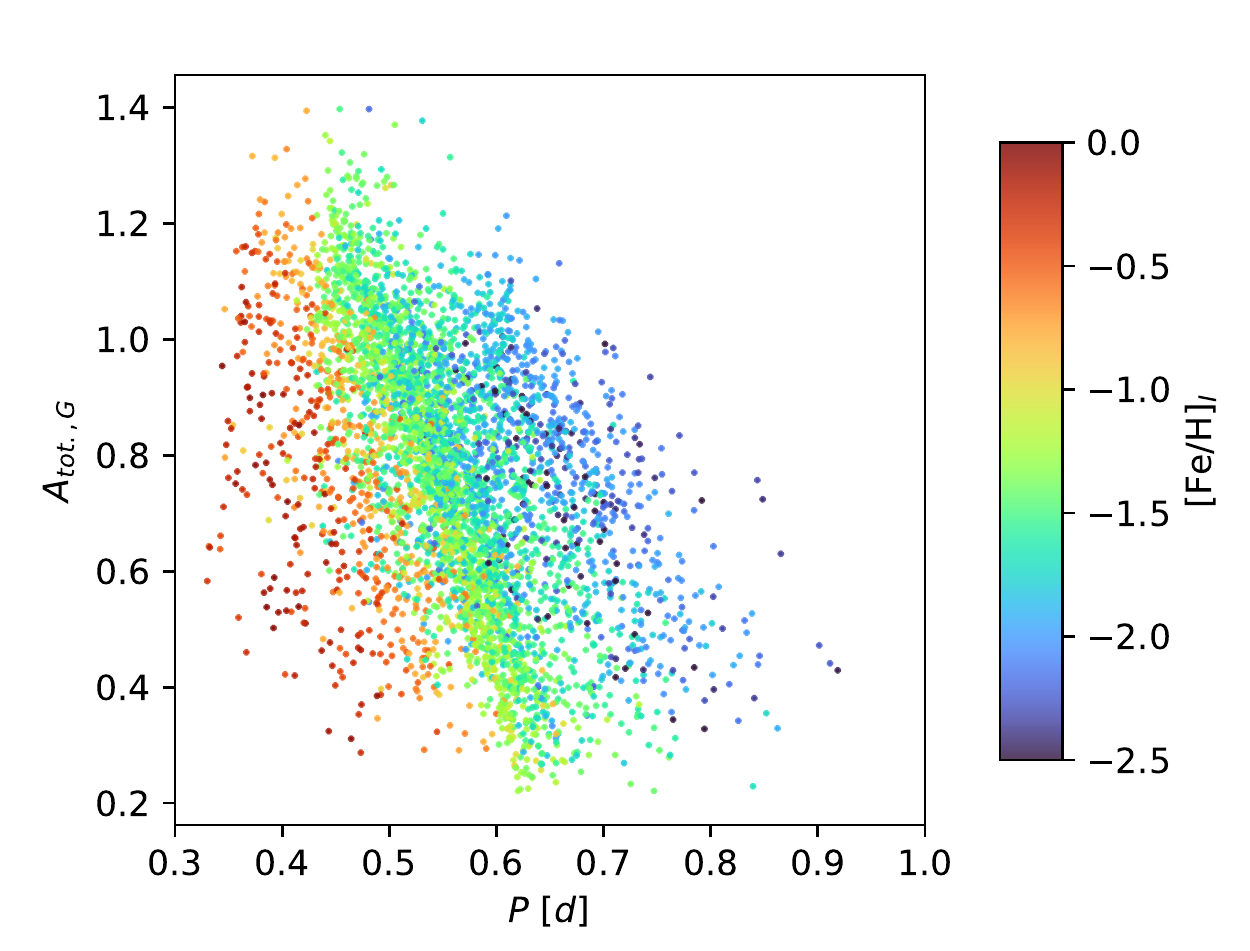}{0.5\textwidth}{}
	}
	\caption{Period--amplitude--metallicity distributions of the VVV $K_s$-band (left) and Gaia $G$-band (right) light curves of the RRab stars in our respective development data sets. The metallicity values were estimated from the $I$-band light curves of the same objects (see text) and are denoted with a color map shown at the right side of the figure.
		\label{fig:per_amp}}
\end{figure*}

By applying the selection criteria in Eqs.~\ref{eq:crit_G} and \ref{eq:crit_Ks}, we remove most Gaia stars from the crowded inner bulge where they have generally poor sampling, and exclude most stars from the relatively small common disk sample of OGLE and VVV, resulting in almost disjunct development data sets for the $G$ and $K_s$ bands. They comprise 4458 $G$-band and 7534 $K_s$-band light curves, which are shown in Fig.~\ref{fig:lightcurves}. The Bailey (period--amplitude) diagrams of the development data sets are shown in Fig.~\ref{fig:per_amp} with the corresponding [Fe/H]$_I$ values color-coded; highlighting the strong correlation of the period on the metallicity, and the more non-linear structure of this diagram in the near-IR compared to the optical domain. 

\begin{figure*}
	\gridline{
		\fig{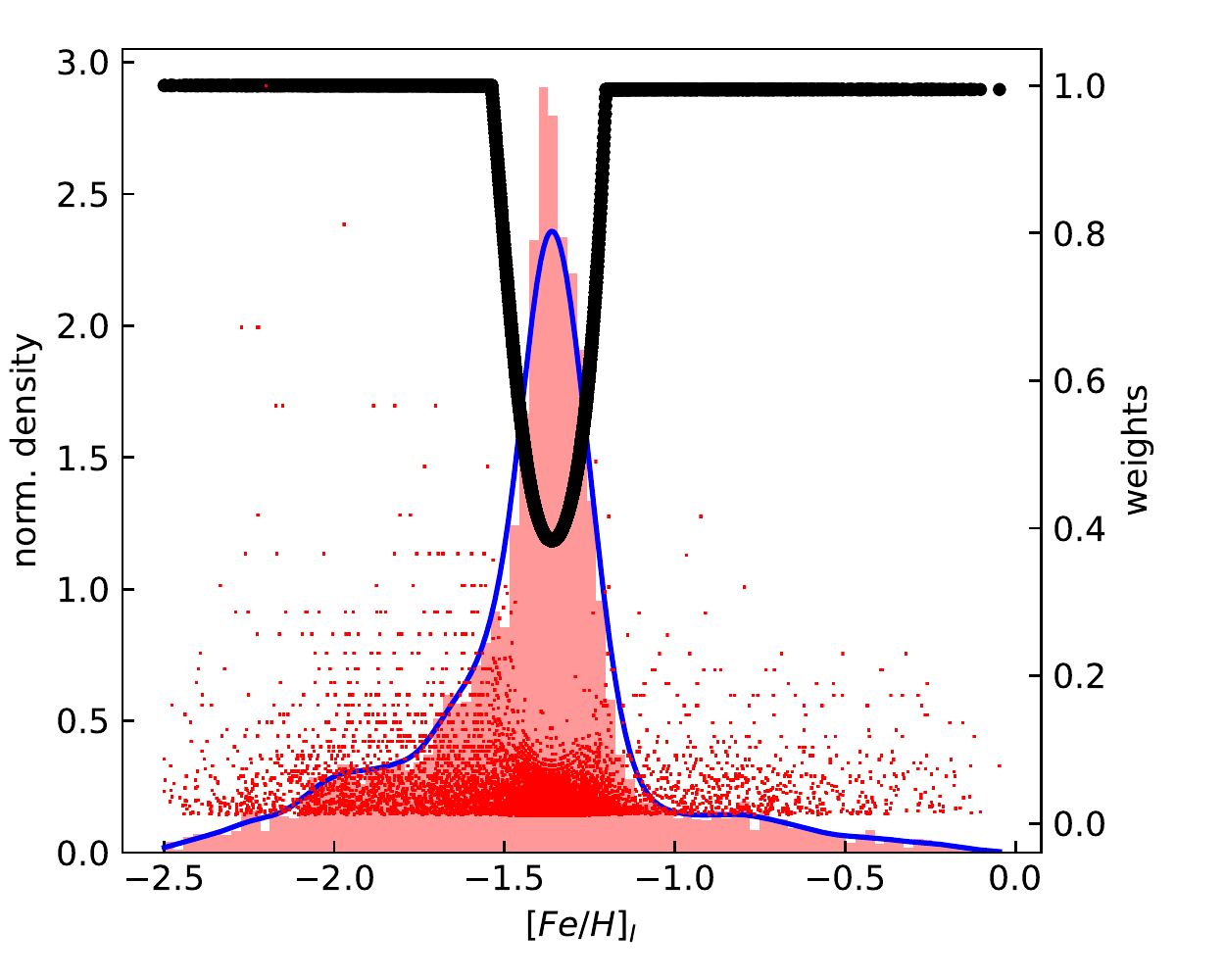}{0.45\textwidth}{}
		\hskip-0.8cm
		\fig{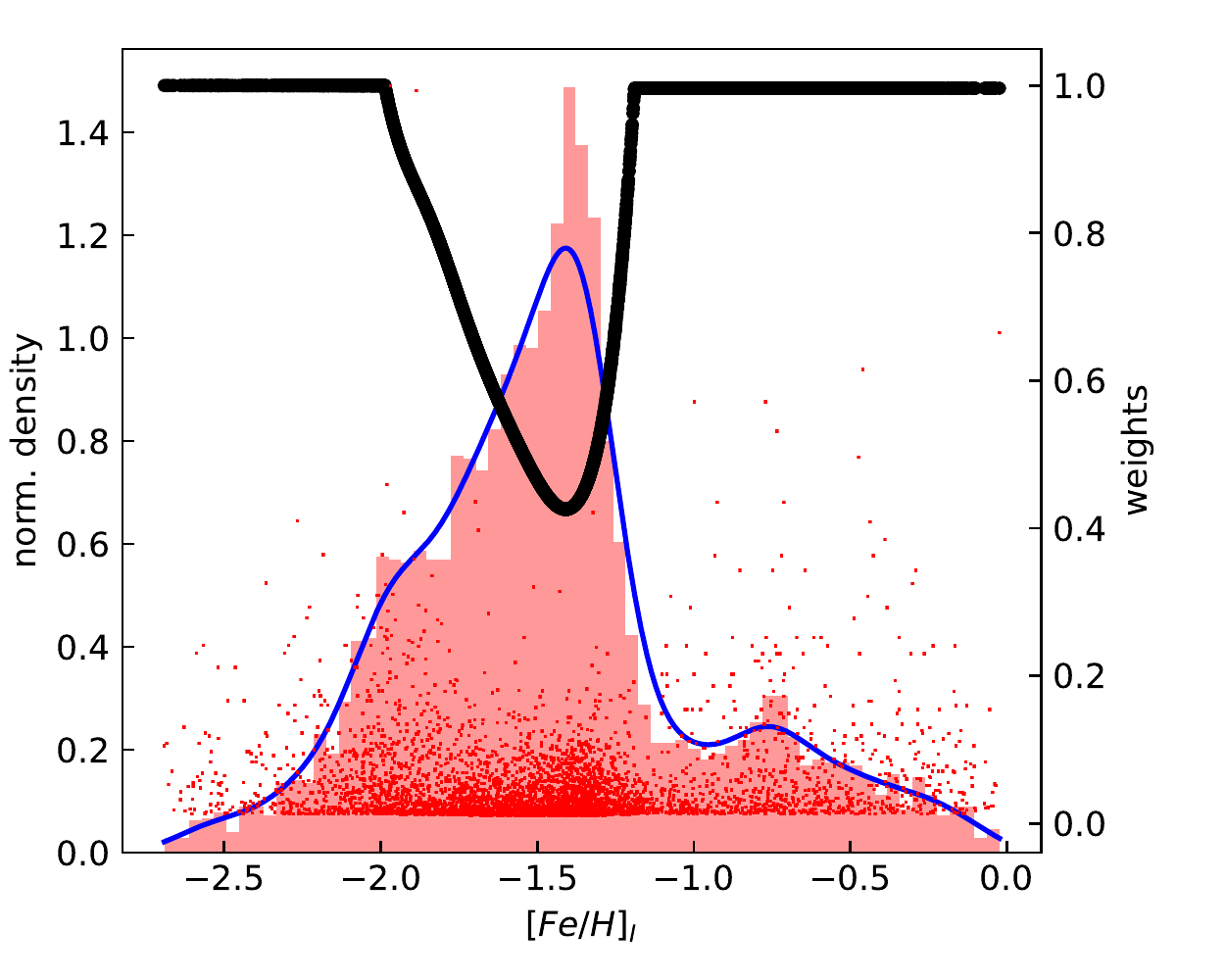}{0.45\textwidth}{}
	}
	\caption{Distributions of the metallicities of the $K_s$- (left) and $G$-band (right) development data sets and their sample weights. Red bars and blue curves show the histograms and kernel density estimates of the [Fe/H]$_I$ values. Black symbols denote the (normalized) weights computed from the inverse of the density. Red points show the final sample weights.
		\label{fig:weighing}}
\end{figure*}

Figure~\ref{fig:weighing} shows the [Fe/H]$_I$ distributions of these data sets, revealing the strong data imbalance, which particularly affects the $K_s$ band. Since both data sets predominantly comprise bulge stars, their metallicity distributions are strongly peaked at around -1.4~dex, with relatively weak tails on both their metal-rich and metal-poor sides.

In order to handle the data imbalance, we introduced density-dependent sample weights for the training of our regression model. First, we computed Gaussian kernel density estimates of the [Fe/H]$_I$ distributions, evaluated them for every object in the development sets, and assigned a density weight $w_d$ to each data point by taking the inverse of the estimated normalized density. Additionally, we introduced density threshold values $\rho_G$ and $\rho_{K_s}$, beyond which uniform $w_d$ values are used in order to prevent an excessive influence of the relatively few data points in the tails of the [Fe/H]$_I$ distributions in the regression. We treat these thresholds as tunable hyper-parameters, and find their optimal values to be $\rho_G = 0.5$ and $\rho_{K_s} = 0.9$. In addition, to take the uncertainties in individual [Fe/H]$_I$ values also into account, we introduce $w_u$ weights as their normalized squared inverse. The final sample weights are then computed as the product $w = w_d \cdot w_u$. Figure~\ref{fig:weighing} shows the distributions of $w_d$ and $w$ for both the $K_s$- and $G$-band development sets.

\section{Predictive modeling}\label{sec:model}

\subsection{Long Short-Term Memory Recurrent Neural Networks}\label{subsect:lstm}

Recurrent neural networks (RNNs) encompass a variety of mathematical models that are capable to accurately approximate extremely complex, highly non-linear interrelations in numerical sequences. They have been successfully employed for supervised learning scenarios in a wide range of research areas involving sequential data, ranging from neural translation to finances \citep[see, e.g.,][for a recent review]{yu_rnn_review}.

The simplest type of RNN takes a sequence $\mathbf{x}^{<t>}$ at its input, and performs the following recurrent transformation at every time step $t\in\{1,T_x\}$:

\begin{equation}\label{RNN}
\mathbf{a}^{<t>} = \tanh(\mathbf{W_{aa}}\mathbf{a}^{<t-1>} + \mathbf{W_{ax}}\mathbf{x}^{<t>}+\mathbf{b_a})~.
\end{equation}

\noindent Here, the output ${a}^{<t>}$ is called the activation of the $t$-th time step, and the $\mathbf{W_{aa}}$ and $\mathbf{W_{ax}}$ weight matrices and $\mathbf{b_a}$ bias vector are free parameters of the RNN, which are shared across all time steps. In case of using the RNN for predicting a single real number from each input sequence, the following operation is performed on the activation vector from the last time step:

\begin{equation}\label{rnn-output}
\hat y  = (\mathbf{w_{ya}}\mathbf{a}^{<T_x>} + b_y)~,
\end{equation}

\noindent where $\hat y$ is the predicted value of the response variable $y$, and the $w_{ya}$ weight vector and the $b_y$ bias term are free parameters of the model.

The RNN's complexity can be adjusted to a specific problem by changing the dimensionality of the weight matrices (a.k.a. the number of neurons) and hence the dimension of the resulting activation vectors, and by using multiple recurrent layers. In case of the latter, the input of layer $l$ will be the sequence of activation vectors $\mathbf{a}^{[l-1]<t>}$ computed by the previous layer $l-1$; each layer will have their own  $\mathbf{W_{aa}}^{[l]}$, $\mathbf{W_{ax}}^{[l]}$ weight matrices and $\mathbf{b_a}^{[l]}$ bias vectors, and the prediction will be computed from the activation vector of the last time step of the last layer by Eq.~\ref{rnn-output}. The optimal model parameters of the RNN are found by minimizing an appropriate cost function, e.g., the mean squared error (MSE) in case of a regression problem.

Modern alternatives to the classical RNN architecture outlined in Eq.~\ref{RNN} include modifications aiming to increase the speed of training and to enable a better representation of long-term dependencies in lengthy input sequences. Perhaps the most successful of these has proven to be the Long Short-Term Memory (LSTM) network originally proposed by \citet{lstm_hochreiter_schmidhuber}, which has been the primary architecture of choice in a wide spectrum of use cases \citep[see, e.g.,][and references herein]{Houdt2020ARO}. Recently, LSTMs have been successfully employed in time-domain astronomy for the classification of photometric time series, e.g. to identify RR~Lyrae stars \citep{dekany_near-infrared_2020} and detect stellar flares \citep{2021A&A...652A.107V}.

A standard LSTM unit features a memory cell designed to retain information over an arbitrary number of time steps, and thus enable the network to learn long-term dependencies. It also contains update, forget, and output gates that regulate the information flow to/from the memory cell and enable to reset its state. The cell and the gates each have free parameters that are learned by the model (i.e., fitted to the data). For a more in-depth mathematical discussion of LSTM networks in the context of astronomical light curves, we refer to \citet[][see their Eqs.~6--11 and Fig.~1]{dekany_near-infrared_2020}.

\subsection{Model selection and optimization}\label{subsec:optimization}

The optimization of a machine-learned predictive model consists of two fundamental steps, training and hyper-parameter tuning. During training, the optimal values of a model's parameters are found by minimizing a cost function for a training data set. Since the gradients of the cost function can be explicitly expressed in the case of neural networks, gradient descent-based optimization algorithms can be employed. The model's hyper-parameters that govern its complexity, such as the number of layers and the number of neurons in each of them, along with the data filtering and weighing thresholds (see Sect.~\ref{sec:data}), as well as the choice of regularization and its parameter(s) have fixed values during training. The optimal values of these hyper-parameters are searched for by maximizing a performance metric for a validation data set that the model has not seen during training. 

For training, we used the MSE cost function with sample weights discussed in Sect.~\ref{sec:data}. In order to prevent overfitting the model to the training set, we experimented with two different methods, namely kernel regularization and dropout. In the former, the $J$ cost function includes a term that is proportional to the norm of weights that are used for the linear transformation of the input sequence and/or the recurrent state (i.e., previous activation vector):

\begin{equation}\label{cost}
J  \propto \sum_{i=1}^{N}(y-\hat y)^2 + \lambda \vert\vert {\bf W} \vert\vert~,
\end{equation}

\noindent where $N$ is the number of training data.
The coefficient $\lambda$ of the weight norm ${\bf W}$ is called the regularization parameter, which, together with the type of norm used, are handled as hyper-parameters of the model. In addition, we also employed the dropout \citep{JMLR:v15:srivastava14a} regularization technique, whereby a random $P_d$ fraction of the activation vectors' elements are randomly dropped (i.e., assigned to 0) during training, with the dropout probability being another hyper-parameter.

The performance metric to be maximized during hyper-parameter tuning should reflect how well the trained model fulfills our expectations on yet unseen data (i.e., the validation set). On the one hand, we would like to minimize our mean prediction errors, which we formulate as maximizing the coefficient of determination, i.e., the $R^2$ score:

\begin{equation}\label{r2}
R^2 = 1- \frac{\sum_i (y_i - \hat y_i)^2}{\sum_i (y_i - \bar y)^2}~,
\end{equation}

\noindent where $\bar y$ is the mean of the response variable in the validation set. The $R^2$ score measures predictive performance relative to the total variation of the response variable. Its best possible value is $1$ (i.e., no prediction errors), and it can take arbitrarily low values. On the other hand, we expect the predicted distribution of the metallicity to be as close to the real distribution as possible. In other words, we want the predictive performance to be homogeneously high across the entire range of metallicities covered by the development set. We formulate this by minimizing the symmetric difference measure between the true and predicted distributions, i.e., their Jensen--Shannon (JS) divergence:

\begin{equation}\label{jsd}
D_{JS} = \frac{1}{2}(D_{KL}(P \vert \vert M) + D_{KL}(Q \vert \vert M))~.
\end{equation}

\noindent Here, $P$ and $Q$ are the compared distributions, $M = (P + Q) / 2$, and $D_{KL}$ is the Kullback-Leibler divergence. The JS divergence is bound to the $[0,1]$ interval and takes the value of $0$ in case the compared distributions are identical. An optimal value of the $R^2$ metric does not necessarily mean that $D_{JS}$ takes its optimum as well, in case the predictive performance changes with [Fe/H]. Since the data imbalance in our regression problem may enhance such a skew in the model's performance, we search for the optimal hyper-parameter values by jointly optimizing $R^2$ and $D_{JS}$, which we achieve by using their ratio as our custom metric:

\begin{equation}\label{metric}
\mathcal{M} = R^2 / D_{JS}
\end{equation}

Since our development data sets have modest sizes, a single held-out validation set would significantly decrease the diversity of the training data. To avoid this, we could apply $k$-fold cross validation (CV) instead, whereby the development set is randomly split into $k$ training and $k$ disjunct validation sets, a mean performance metric is computed from the predictions on the latter; and the model with the best set of hyper-parameters is finally refitted to the entire development set in order to obtain a single ``monolithic'' final model. Due to the limited amount of data however, this approach can still be volatile to the distribution changes introduced by the training--validation splits. Firstly, the same hyper-parameters can lead to significant over- or underfitting in different folds. Secondly, the true performance of the refitted model can be significantly different from the performance estimate from CV due to the reduced data set sizes in the latter. It might even be the case that the optimal hyper-parameters are different for the final fit to the development set from the ones found by CV.

A more robust predictive model can be obtained by a modified approach to $k$-fold CV. In order to prevent the model's volatility to data splitting during each CV fold, we apply the ``early stopping'' regularization technique, which is commonly used in deep learning scenarios. During training, we monitor the model's performance by evaluating it on both the current training and validation folds after each training epoch. When overfitting occurs, the model's performance measured on the training set will keep increasing as the learning algorithm starts to fit the model to the noise pattern of the training set, while the thus far increasing validation performance will turn over and start to decrease, due to the model's poor generalization. We obtain the optimal model for each CV fold by stopping its training at this turnover point, and estimate its overall performance by the mean $\mathcal{M}$ metric across the folds. Finally, instead of refitting the model with the best hyper-parameters to the development set as we would in standard $k$-fold CV, we instead keep the $k$ models and use their mean prediction. This approach is commonly referred to as training an ``early stopped $k$-fold ensemble'' of models, and is not only more robust than a single monolithic model, but offers the advantage that instead of providing a point estimate for an input, its prediction's uncertainty can also be estimated as the standard deviation of the ensemble's $k$ predictions.

An extensive grid search of hyper-parameters was performed to find the best model ensemble. We tried one- and two-layer mono- and bidirectional LSTM (biLSTM) architectures with various numbers of neurons in each. The effect of both L1 and L2 norms were tested in Eq.~\ref{cost} for the cost function's weight penalty (a.k.a. lasso and Tikhonov regularization, respectively), along with a grid of $\lambda$ values and dropout probabilities. We trained each model with the Adam optimization algorithm \citep{2014arXiv1412.6980K} with a learning rate of $0.005$ until the optimal early stopping epoch was found (typically after a few thousand epochs). In order to have a sufficient number of training examples from our entire [Fe/H] range, we used a large mini-batch size of $256$\footnote{Each training iteration is performed on a mini-batch of training data. During one training epoch, all training data are used once.}. 

\begin{figure}
	\gridline{
		\fig{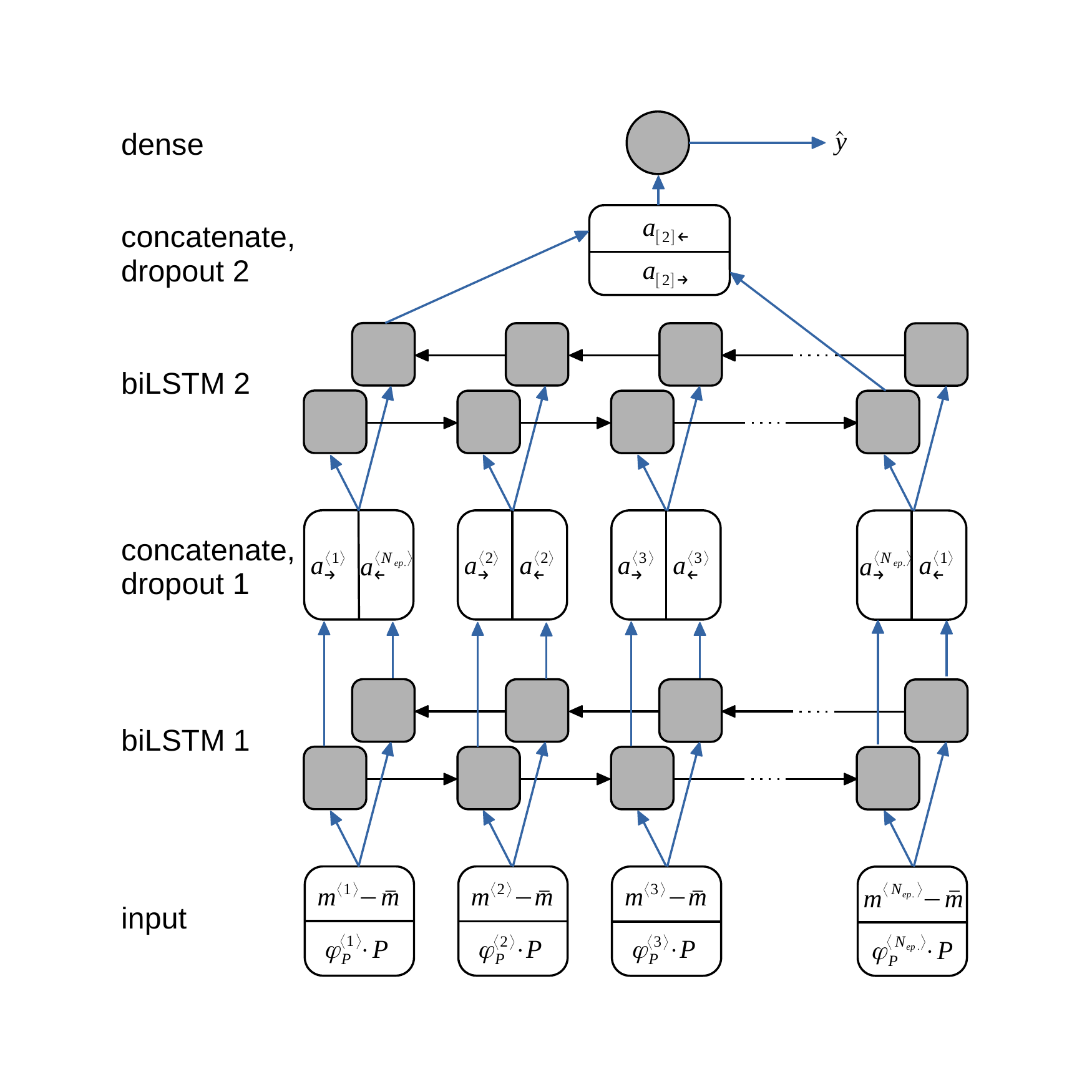}{0.55\textwidth}{}
	}
	\caption{Schematic architectural graph of our predictive models. Grey boxes represent neural network (LSTM and dense) units, empty boxes denote their inputs and outputs. The activation vector at the output of the first biLSTM layer is 64 and 32 dimensional for the $K_s$- and $G$-band model, respectively.
		\label{fig:network}}
\end{figure}

Figure~\ref{fig:network} shows a schematic picture of our best-performing predictive models, while their main attributes are listed in Table~\ref{tab:best_models}. For both the $K_s$ and $G$ bands, they contain two biLSTM layers, and use L1 regularization on the recurrent weights. In addition, dropout is employed after both the first and second layers.

\begin{deluxetable*}{lllll}
	\tablecaption{Properties of the final $K_s$- and $G$-band predictive models of the metallicity\label{tab:best_models}}
	\tablehead{
		 & \multicolumn2c{$K_s$-band model} & \multicolumn2c{$G$-band model} \\
		\colhead{Layer} & \colhead{hyper-params.} & \colhead{params.} & \colhead{hyper-params.} & \colhead{params.}
	}
	\startdata
	biLSTM1  & 32 neurons, $\lambda=3\cdot10^{-6}$ & 8960 & 16 neurons, $\lambda=5\cdot10^{-6}$ & 2432 \\
	dropout1 & $P_d = 0.1$            & 0 & $P_d = 0.1$      & 0 \\
	biLSTM2  & 32 neurons, $\lambda=3\cdot10^{-6}$ & 24832 & 16 neurons, $\lambda=5\cdot10^{-6}$ & 6272 \\
	dropout2 & $P_d = 0.1$            & 0 & $P_d = 0.1$      & 0 \\
	dense    & --               & 65 & --   & 33 \\
	\enddata
	\tablecomments{The number of neurons refers to a single direction of the network.}

\end{deluxetable*}

\subsection{Regression performance}\label{subsec:performance}

The regression performance of our final model ensembles has been measured using various common metrics applied on the union of the $k$-fold validation data sets, except in the case of the $R^2$ metric, for which the mean was computed. All of these metrics estimate very low generalization error, i.e., high prediction performance on unseen data. Their values for both the $K_s$ and the $G$ bands are shown in Table~\ref{tab:metrics}, which also includes the same metrics for the training data as reference. The similarity between the values obtained for the training and validation data sets indicates an excellent bias-variance tradeoff of the final model ensembles. We emphasize that the metrics in Table~\ref{tab:metrics} describe our model ensembles' precision in predicting the D21 $I$-band photometric metallicities. Their accuracy in predicting the [Fe/H] can be estimated by combining the metrics in Table~\ref{tab:metrics} with those of the D21 formula. For example, by adding the MAE values in quadrature, we obtain a total nominal uncertainty of $\sim0.19$~dex.

\begin{deluxetable}{lllll}
	\tablecaption{Various performance metrics of our final LSTM model ensembles 
	in terms of predicting the $I$-band photometric metallicities}
	\tablehead{
		& \multicolumn2c{$K_s$-band model} & \multicolumn2c{$G$-band model} \\
		\colhead{Metric} & \colhead{tr.} & \colhead{val.} & \colhead{tr.} & \colhead{val.}
	}
	\startdata
	$R^2$			& 0.87 & 0.84 & 0.96 & 0.93 \\
	wRMSE			& 0.12 & 0.13 & 0.10 & 0.13 \\
	wMAE			& 0.09 & 0.10 & 0.07 & 0.10 \\
	RMSE			& 0.14 & 0.15 & 0.15 & 0.18 \\
	MAE				& 0.10 & 0.11 & 0.12 & 0.13 \\
	medAE			& 0.07 & 0.08 & 0.09 & 0.10 \\
	$D_{JS}$		& 0.004& 0.004& 0.001& 0.001\\
	$R^2/D_{JS}$	& 218  & 239  & 956  & 960  \\
	\enddata
	\tablecomments{(w)RMSE: (weighted) root mean squared error; (w)MAE: (weighted) mean absolute error; medAE: median absolute error; $D_{JS}$: Jensen-Shannon divergence}
	
\end{deluxetable}\label{tab:metrics}

\begin{figure*}
	\gridline{
		\fig{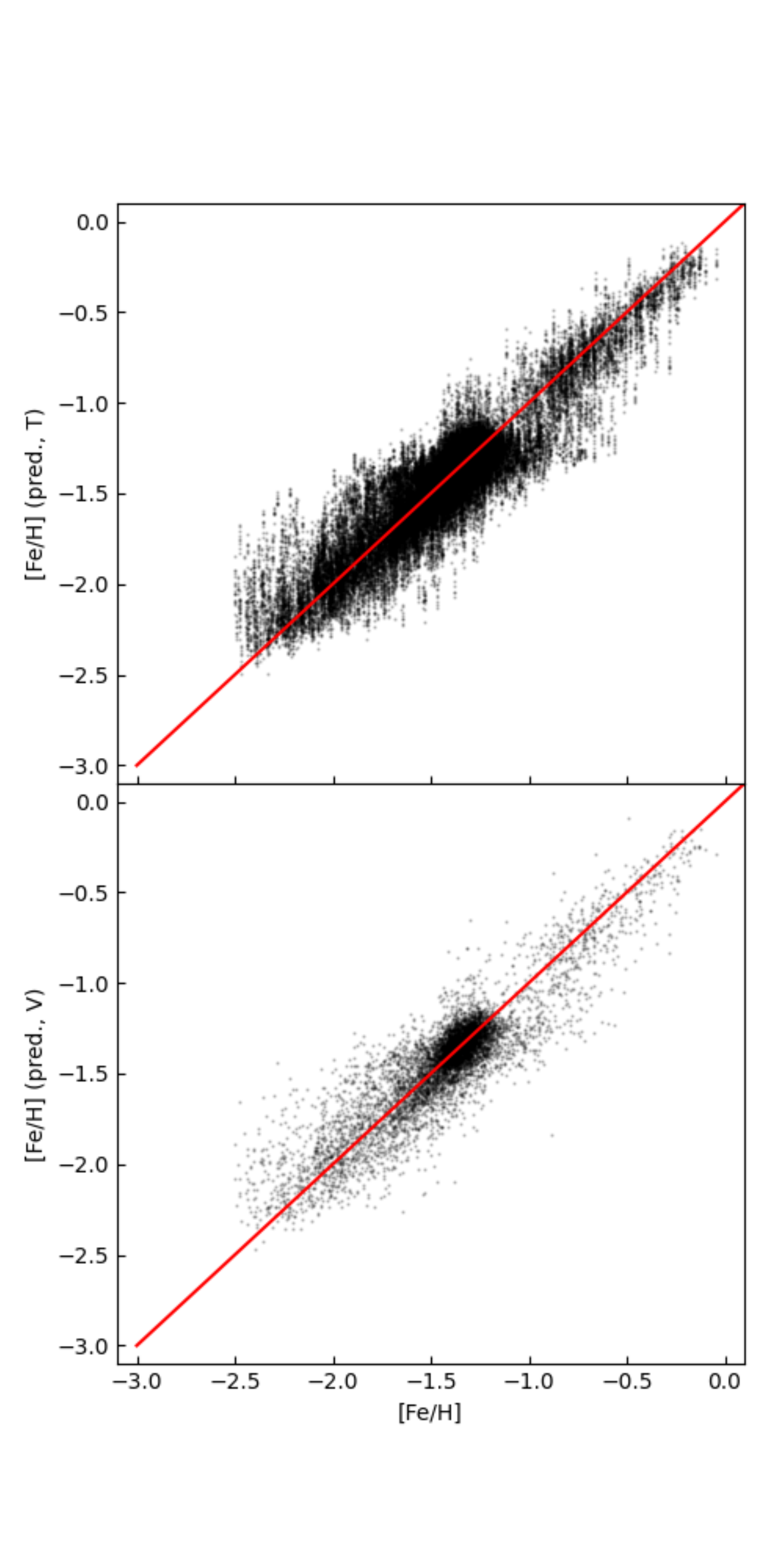}{0.4\textwidth}{}
		\hskip-1cm
		\fig{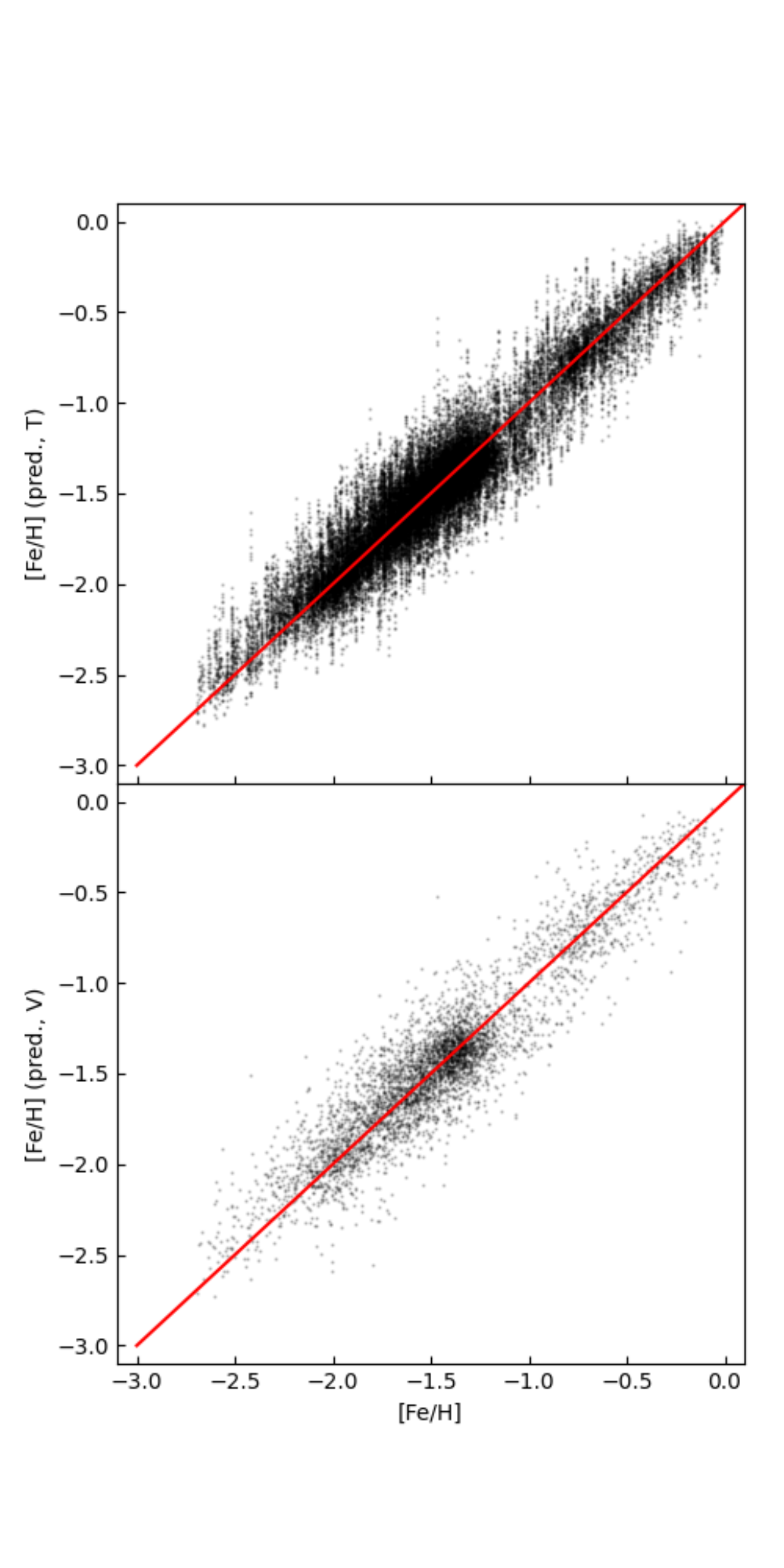}{0.4\textwidth}{}
	}
	\caption{Predicted {\em vs} true photometric metallicities from the best-performing predictive models for the $K_s$ (left) and $G$ (right) wavebands. The top and bottom panels show the full training and validation data sets, respectively. The red lines denote the identity function.
		\label{fig:pred_vs_true}}
\end{figure*}

Figure~\ref{fig:pred_vs_true} compares our model ensembles' [Fe/H] predictions with the metallicities computed from the corresponding $I$-band light-curves. The residuals do not show any significant structure and contain only few strong outliers. At the very metal-poor end ($[{\rm Fe/H}]<-2$), some $K_s$-band metallicity predictions lie out from the main locus with a sizable positive bias for a relatively small number of stars. Upon the inspection of both their $K_s$- and $I$-band light curves, we did not identify peculiarity in their photometry. It might be possible that a latent physical parameter with a large influence on the light-curve shape (such as the He content) is systematically different for these objects compared to the rest of the sample. Another possibility is that the dependency of the $K_s$ light-curve shape on the [Fe/H] is too complicated at the metal-poor end for our model to accurately learn it from the available data.

\begin{figure*}
	\gridline{
		\fig{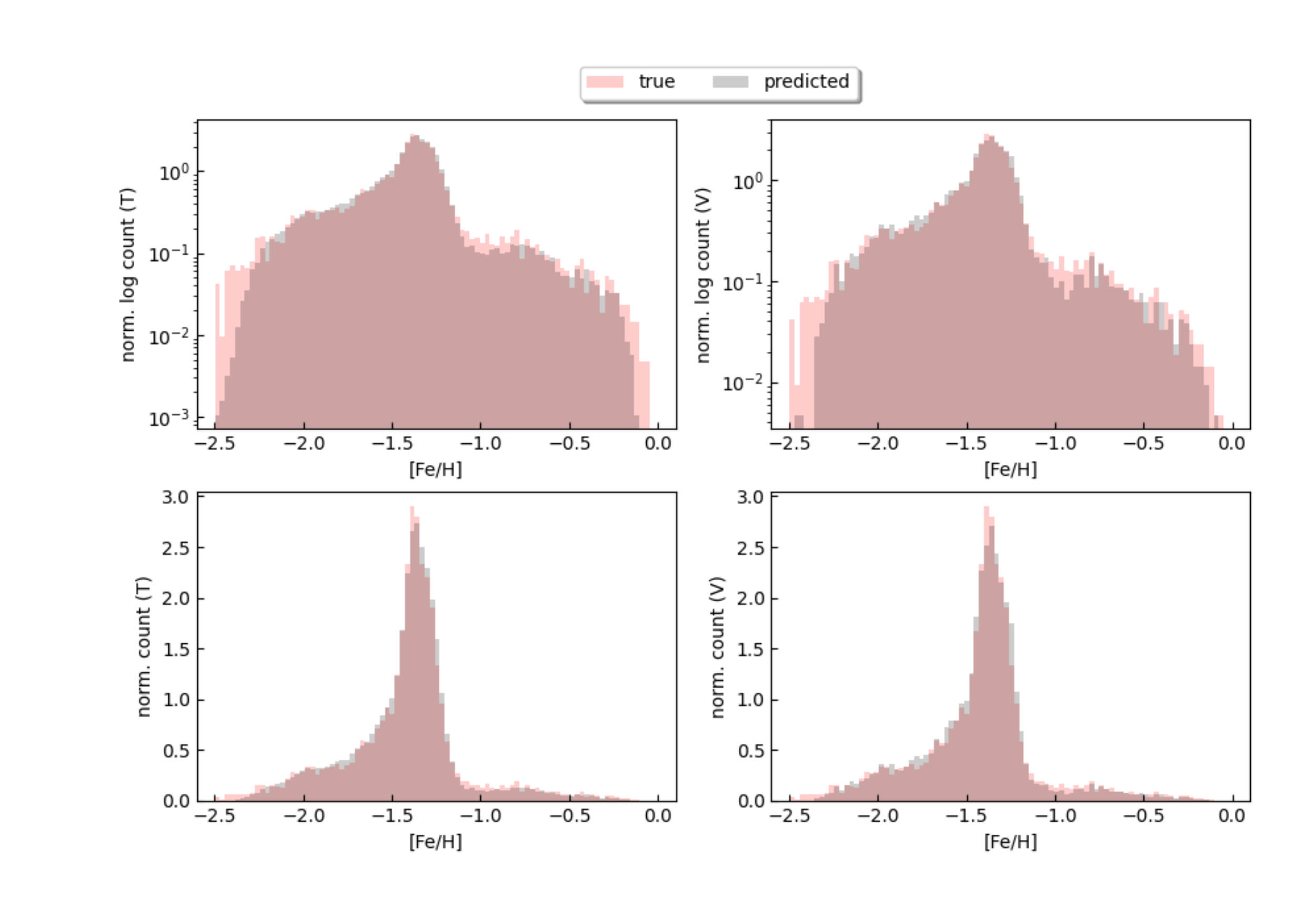}{0.49\textwidth}{$K_s$ band}
		\hskip-1cm
		\fig{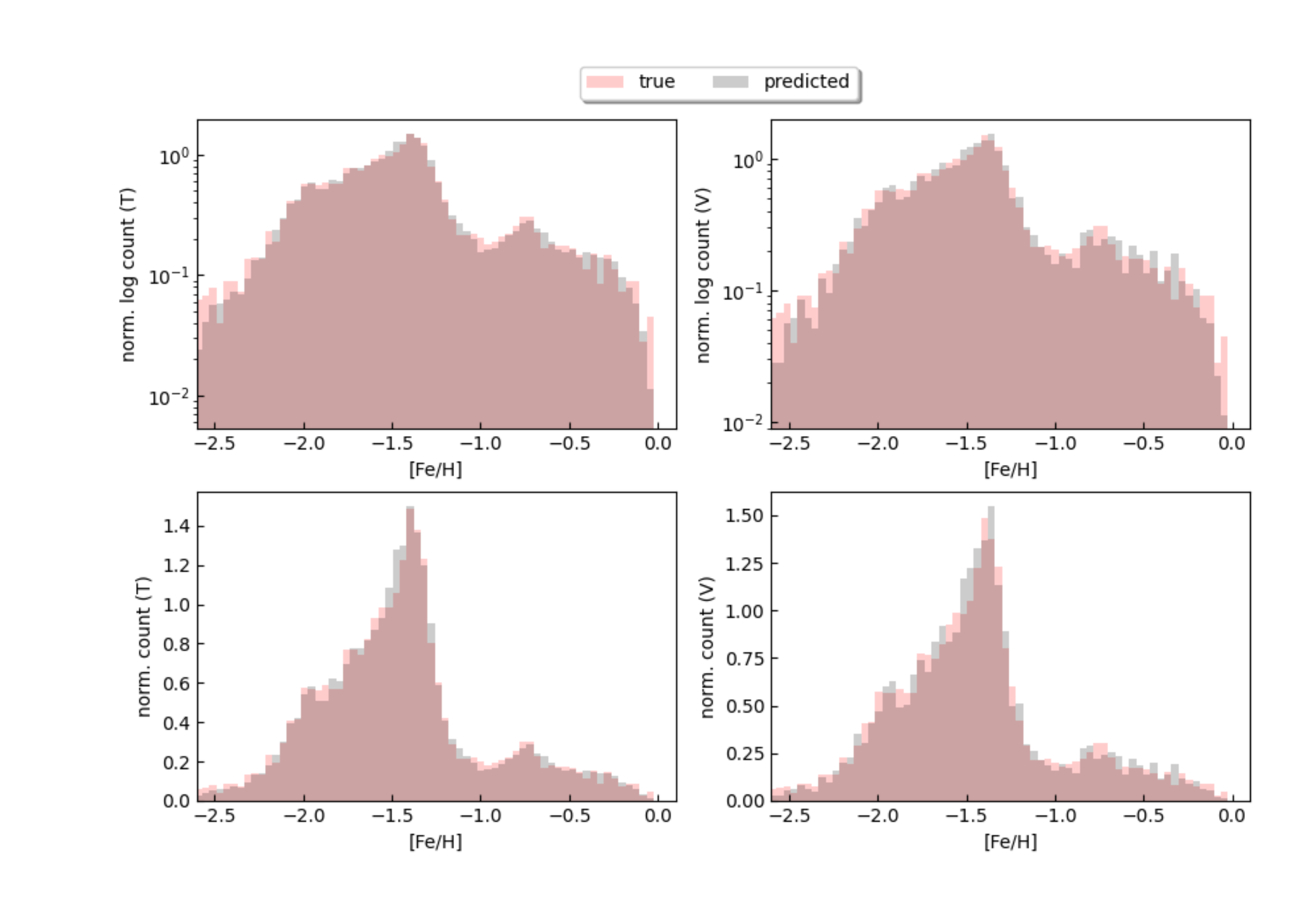}{0.49\textwidth}{$G$ band}
	}
	\caption{Histograms of the true (red) and predicted (gray) [Fe/H] values from our $K_s$- and $G$-band model ensembles for the training (T, left) and validation (V, right) data sets. The upper and lower panels show the histograms on logarithmic and linear scales, respectively.
		\label{fig:pred_true_hist}}
\end{figure*}

The photometric metallicity distributions predicted by our model ensembles for the training and validation data sets are directly compared to the $I$-band photometric metallicities in Fig.~\ref{fig:pred_true_hist}. For both wavebands, the predicted distributions are virtually identical to the original ones, not showing any bias, and even subtle features are accurately recovered, which tallies with their very low $D_{JS}$ values.

\begin{figure}
	\gridline{
		\fig{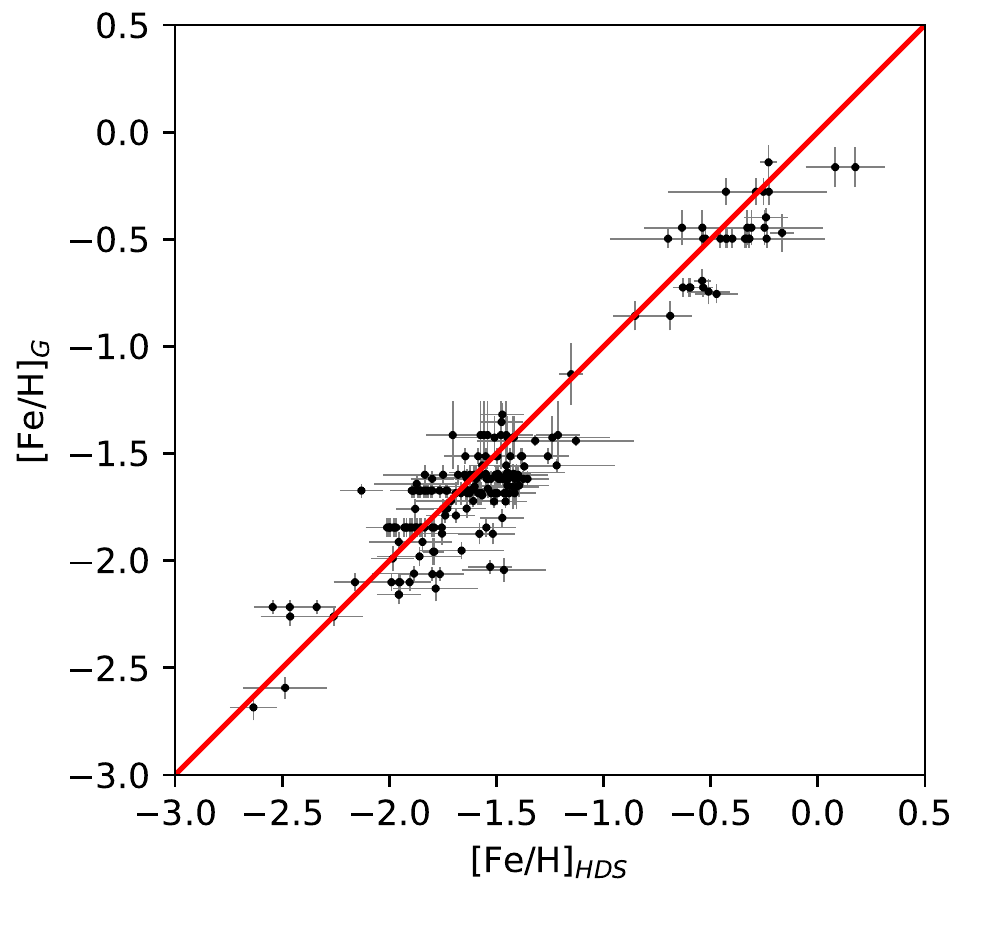}{0.4\textwidth}{}
	}
	\caption{$G$-band mean [Fe/H] predictions of 60 RRab stars by our LSTM model ensemble plotted against the individual HDS metallicity measurements of the same objects. The red line denotes the identity function.
		\label{fig:hrs_vs_gfeh}}
\end{figure}

We further test the consistency of our models by directly comparing their [Fe/H] predictions with HDS measurements of  RRab stars in the Gaia DR2 catalog. We cross-matched the latter with the same large spectroscopic sample that was compiled from the literature for the development of our D21 $I$-band photometric metallicity estimator, using the results of 
\citet[C21]{crestani_deltaS},
\citet[F11]{for_chemical_2011}, \citet[C17]{chadid_spectroscopic_2017}, 
\citet[S17]{sneden_rrc_2017},
\citet{clementini_composition_1995}, \citet{fernley_metal_1996},
\citet{lambert_chemical_1996}, 
\citet{liu_abundances_2013}, 
\citet{nemec_metal_2013}, 
\citet{govea_chemical_2014}, 
\citet{pancino_chemical_2015}, and \citet{andrievsky_relationship_2018}. The corrections for the systematic [Fe/H] offsets computed by D21 were applied to these measurements to match the homogeneous scale of C21+F11+C17+S17. We selected the objects with well-sampled and accurate $G$-band light curves, which resulted 213 individual data points of 60 unique objects, 28 with single and 32 with multiple HDS [Fe/H] measurements. We note that this data set largely overlaps the one used for calibrating the D21 predictive formula, but is too small and imbalanced for a reliable direct calibration of a similar formula for the $G$ band. Figure~\ref{fig:hrs_vs_gfeh} compares the predicted [Fe/H]$_G$ values with the corresponding individual HDS measurements. The two sets of values are in good agreement, showing no systematics in their residual, and their scatter is consistent with the prediction uncertainties measured on the validation data set (c.f. Fig~\ref{fig:pred_vs_true}). This important cross-check indicates the pertinence of our underlying assumption about the physical similarity between the HDS training set of the D21 formula and the development data set of our LSTM model, thus supporting the consistency of our approach.

\section{Photometric metallicity catalogs}\label{sec:catalogs}

As the final goal of this study, we computed photometric metallicities for a large number of RRab stars by deploying the neural networks developed in Sect.~\ref{sec:model} on the photometric databases of the Gaia and VVV surveys.

\subsection{$G$-band photometric metallicities of the Gaia DR2 RRab stars}\label{subsec:gfeh}

We analyzed the $G$-band light curves of all 98024 RRab stars in the Gaia DR2 RR~Lyrae catalog following the same procedure that was applied on the development set, as described in Sect.~\ref{sec:data}. We adopted the periods and type classifications from \citet{clementini_gaia_2019}. Our predictive model was applied to all light curves that passed the following quality criteria: $C>0.85$; $S/N>30; ~N_{ep.}>20; A_{tot.}<1.4$, resulting in a {\em target set} of 58652 RRab stars. Their metallicity estimates, along with various photometric attributes, are presented in Table~\ref{tab:gfeh}. The sample is dominated by halo stars, with additional large samples contributing to it from the Magellanic Clouds, the bulge, and the Milky Way's thick disk.

In our D21 study, we have shown that several earlier methods for metallicity estimation are affected by systematic positive biases with respect to our new $I$-band [Fe/H] prediction formula, which is directly calibrated to modern HDS measurements. Since the neural networks in this study were trained on $I$-band photometric metallicities obtained with the latter, we can expect to see similar differences between our model predictions and other $G$-band [Fe/H] estimates from the literature.

\begin{figure*}
	\gridline{
		\fig{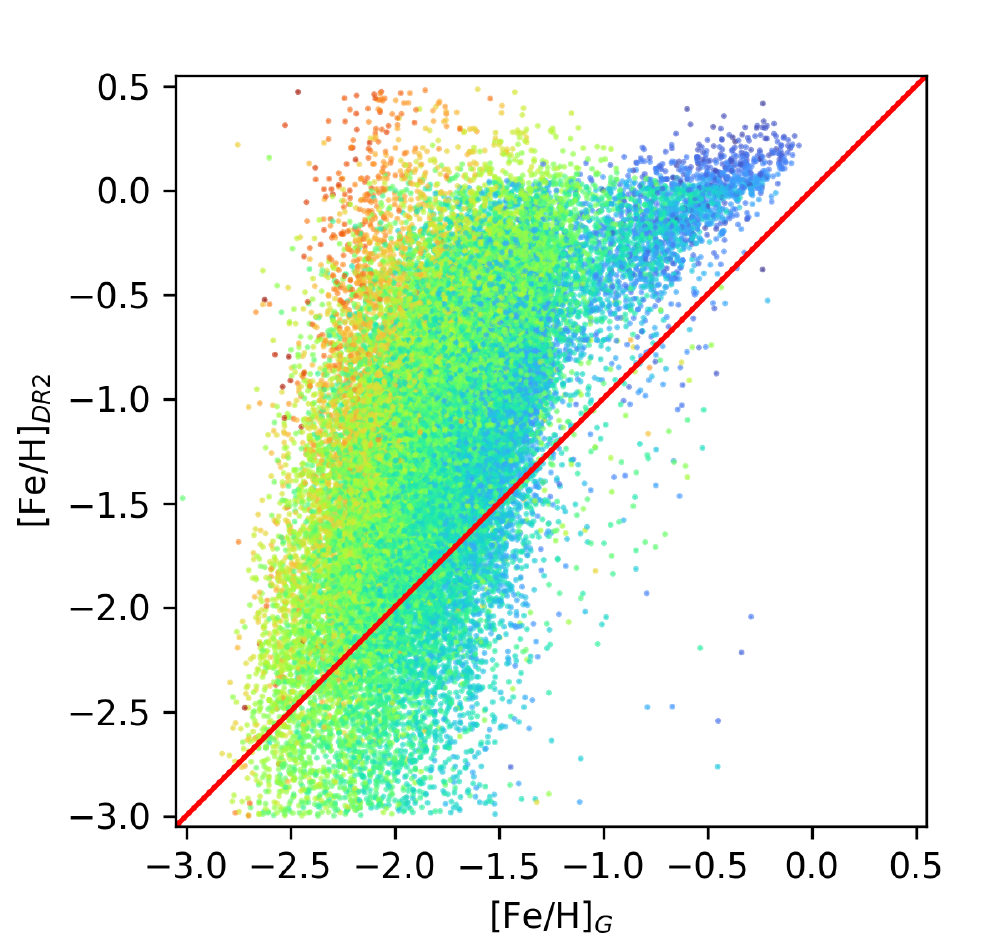}{0.3\textwidth}{}
		\fig{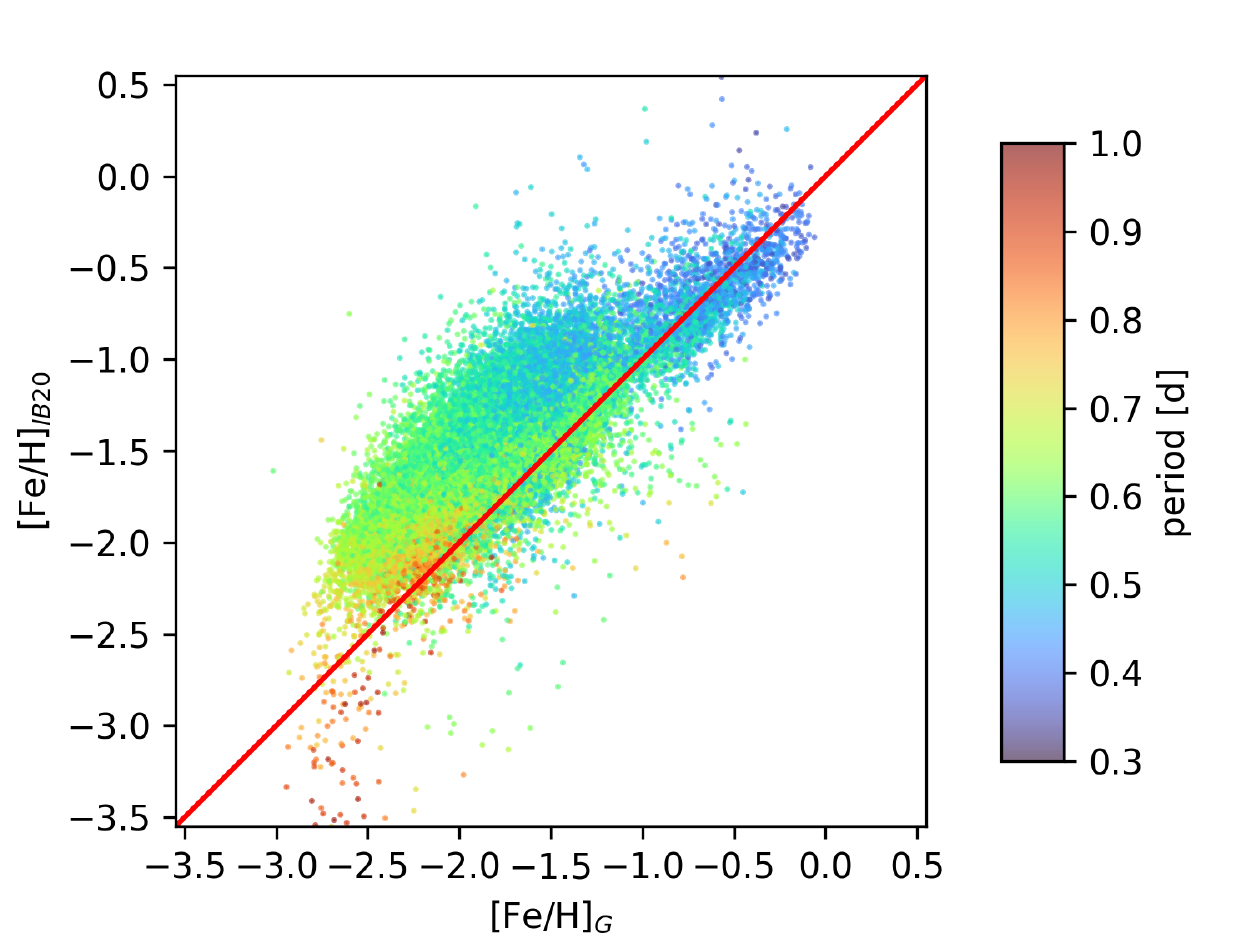}{0.375\textwidth}{}
		\fig{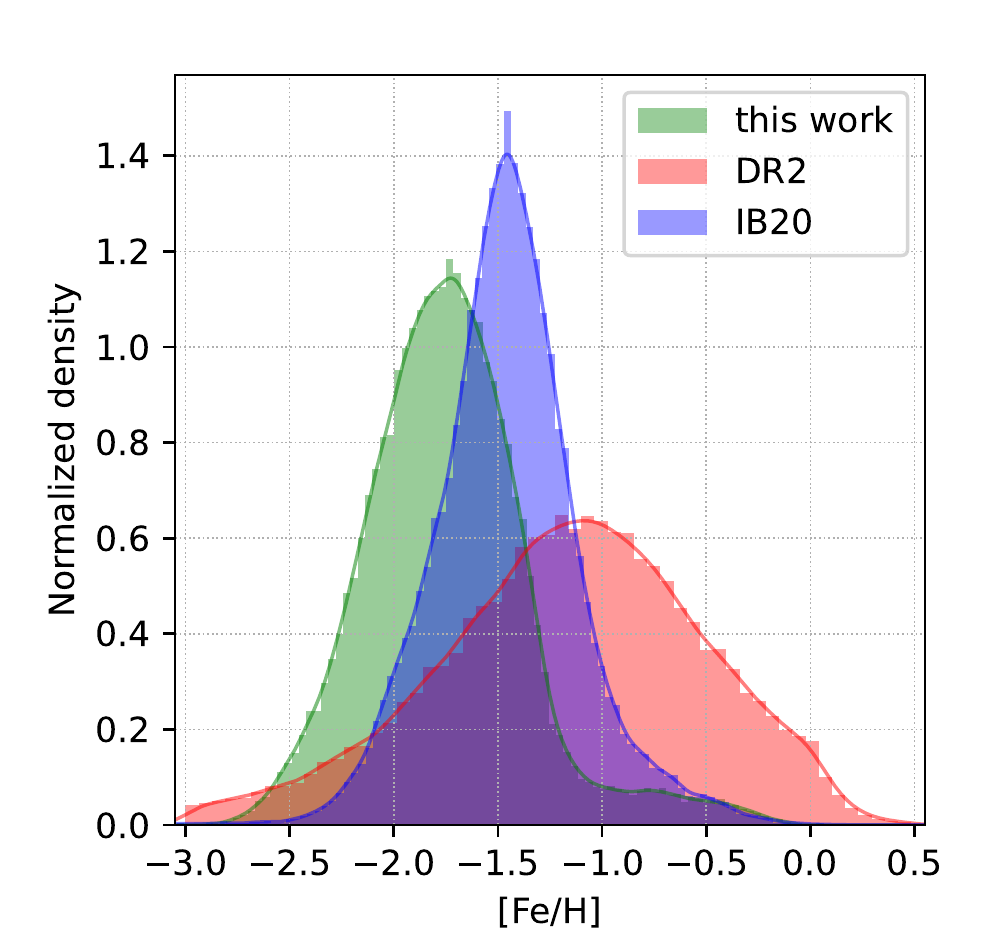}{0.3\textwidth}{}
	}
	\caption{Photometric metallicities predicted by our model ensemble ([Fe/H]$_G$) for the $G$-band target data set, plotted against the corresponding metallicity estimates from the Gaia DR2 RR~Lyrae catalog ([Fe/H]$_{\rm DR2}$, left), and by the IB20 formula ([Fe/H]$_{\rm IB20}$, middle). The pulsation periods are color-coded, the red line denotes the identity function. Right: histograms and corresponding kernel density estimates of the same data set as in the left and middle panels. Green: [Fe/H]$_G$ estimates from this work, blue: IB20 metallicities, red: [Fe/H] values from the DR2 RR~Lyrae catalog. 
		\label{fig:gfeh_dr2feh_ib20feh}}
\end{figure*}

We directly compared our results to metallicities published in the Gaia DR2 RR~Lyrae catalog by \citet{clementini_gaia_2019} ([Fe/H]$_{DR2}$), which were available for 43681 objects in our target set. These estimates are based on the quadratic formula by \citet{nemec_metal_2013}, which relates the metallicity to the period and the $\phi_{31}$ Fourier parameter in the {\em Kepler} photometric waveband, and was calibrated on a tiny sample of 27 HDS measurements. To apply it to $G$-band light-curve parameters, \citet{clementini_gaia_2019} performed two successive linear transformations of the $G$-band $\phi_{31}$ parameters, converting them first to the $V$ band, and from there to the {\em Kepler} band. The left panel of Fig~\ref{fig:gfeh_dr2feh_ib20feh} confronts the [Fe/H]$_G$ values from our study with their [Fe/H]$_{DR2}$ equivalents. The latter are affected by a very large positive bias, and the residual strongly correlates with the period, with long-periodic stars having extreme biases of up to $\sim2$~dex. In addition, the distribution shows a clump of metal-rich stars standing out from the main locus that correspond to a metal-rich tail of the [Fe/H]$_G$ distribution, but remains blended with the rest of the [Fe/H]$_{DR2}$ sample due to its aforementioned bias.

To improve the quality of RR~Lyrae metallicity estimation from Gaia photometry, \citet[][IB20]{iorio_chemo-kinematics_2020} fitted a linear relation to the pulsation period and the $G$-band $\phi_{31}$ Fourier parameter using 84 stars with [Fe/H] values adopted from \citet{layden_metallicities_1994}. The latter were derived from low-resolution spectral indices using a linear formula that was calibrated to early HDS measurements of a handful of field and cluster RRab stars. We note that the same \citet{layden_metallicities_1994} sample, with slight adjustments, forms partly the basis of the widely used photometric metallicity prediction formulae of \citet{jurcsik_determination_1996} and \citet{smolec_metallicity_2005}. In the middle panel of Fig.~\ref{fig:gfeh_dr2feh_ib20feh}, we compare the IB20 metallicities with our [Fe/H]$_G$ estimates for the same set of stars as for our previous comparison. They show a much better agreement with our results than the [Fe/H]$_{DR2}$ values, especially at the metal-rich end, and instead of the huge positive bias observed in the case of DR2 metallicities of long-periodic, metal-poor stars, we see a smaller, but still quite large bias in the opposite, negative direction. Overall, the IB20 formula also overestimates the metallicities with respect to our model, with a positive bias of up to 0.2--0.3~dex at the center of the distribution. The particularly poor DR2 and IB20 predictions at the long-periodic, metal-poor limit can probably be attributed mainly to the lack of such stars in the data sets used to calibrate those formulae.

Finally, we compare the photometric metallicity distributions using all three methods in the right panel of Fig.~\ref{fig:gfeh_dr2feh_ib20feh}, revealing large systematic differences between them. In addition to large offsets between their centers, their shapes are also remarkably different. Our distribution features a long metal-rich tail, which is absent from the DR2 MDF, and only slightly apparent in the IB20 MDF, because it blends with the bulk of the distributions due to the positive bias dominating the latter.

\begin{figure*}
	\gridline{
		\fig{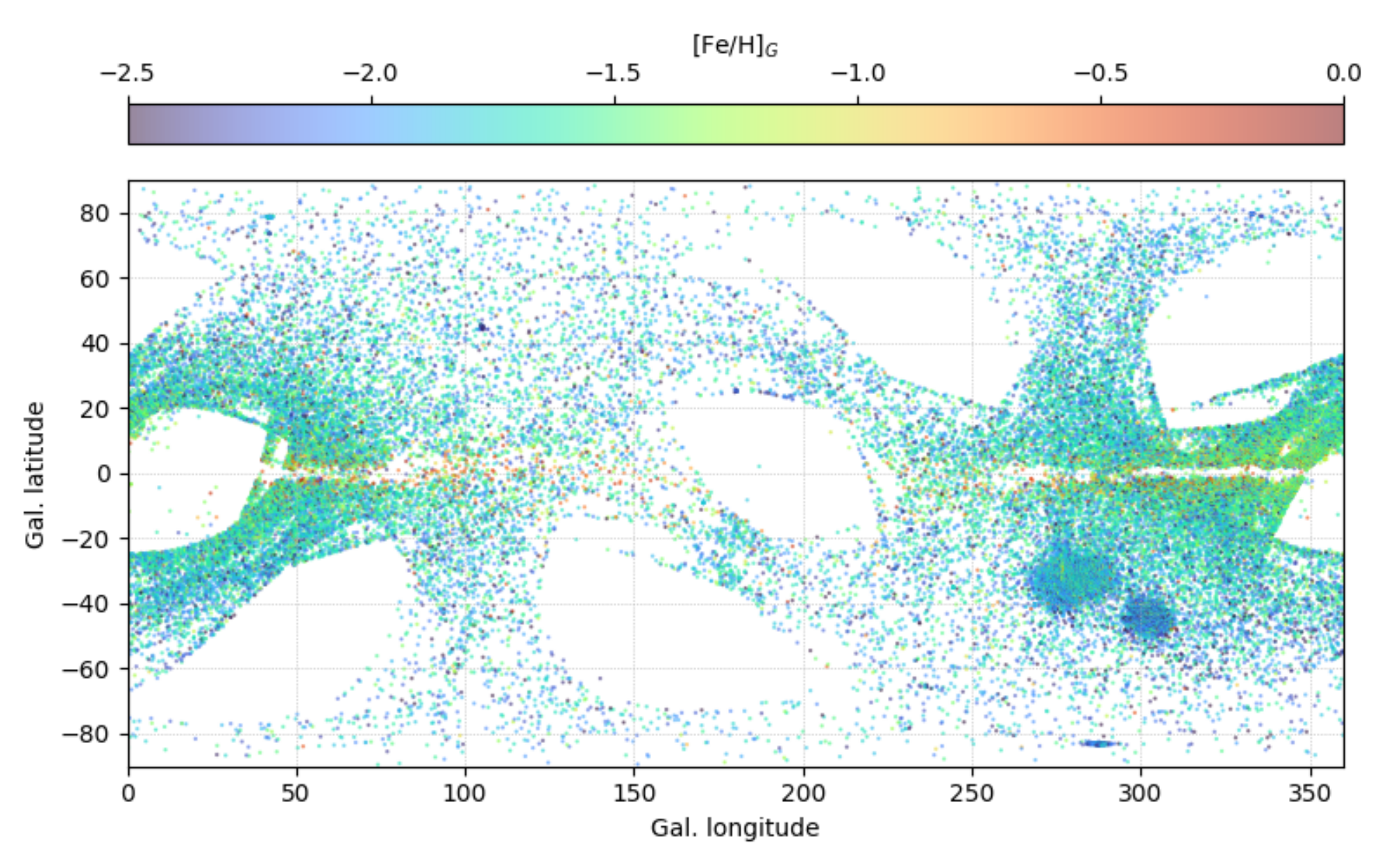}{0.8\textwidth}{}
	}
	\caption{Celestial distribution of the target data set of our Gaia $G$-band photometric metallicity estimator in the Galactic coordinate system. The predicted [Fe/H]$_G$ values are color-coded.
		\label{fig:gfeh_map}}
\end{figure*}

Figure~\ref{fig:gfeh_map} shows the celestial distribution of our target set, with the predicted [Fe/H]$_G$ values color-coded. The stars in the metal-rich tail of the MDF lie in close proximity of the Galactic plane, while the high-latitude halo sample and the Magellanic Clouds contribute with the most metal-poor stars. A positive gradient in the mean metallicity from the halo towards the center of the bulge is also apparent. Notably, the metallicities of the stars constituting the metal-rich tail are the ones that show the best correlation between our predictions and those from both DR2 and IB20 (albeit with an offset in the former case; c.f. the left and middle panels of Fig.~\ref{fig:gfeh_dr2feh_ib20feh}). This is probably due to the fact that the training data sets used for fitting the latter two relations comprise similar objects from the Solar neighborhood. 

The distributions in Fig.~\ref{fig:gfeh_dr2feh_ib20feh}, and thus their modes are dominated by halo objects. To obtain simple estimates of the median halo metallicity, we cross-matched our sample with the RR~Lyrae stars in the bulge and Magellanic Clouds listed in the OCVS, and excluded all matches within $1''$, resulting in a reduced sample of $\sim33000$ objects. Since the OCVS is a highly complete catalog, this simple approach leaves us with only a small contamination of disk RRab stars, which does not significantly affect the accuracy in terms of estimating the mode of the halo RRab MDF. To measure the latter, we computed kernel density estimates of the MDFs and determined their maxima. Our estimates of the median [Fe/H] of the halo obtained in this way is $-1.67$~dex using our predictive model, $-1.43$~dex using the IB20 formula, and $-1.04$~dex according to the DR2 RR~Lyrae catalog. Our results are in good agreement with pertinent studies of the inner halo MDF using main-sequence stars by \citet{2013ApJ...763...65A} and \citet{2020MNRAS.492.4986Y}, while the IB20 and DR2 values both show significant positive bias.

\subsection{$K_s$-band photometric metallicities of the VVV inner bulge RRab stars}\label{subsec:kfeh}

The $K_s$-band predictive model ensemble was deployed on the catalog of RRab stars in the inner bulge area of the VVV survey discovered by \citet[][D20]{dekany_near-infrared_2020}. These objects have not been detected by the OGLE survey covering the same area due to extremely high interstellar extinction. Our target data set comprises the public $K_s$-band light curves of all 4446 RRab stars in the D20 catalog, processed according to Sect.~\ref{sec:data}. The resulting photometric parameters and metallicity estimates are displayed in Table~\ref{tab:kfeh}.

\begin{figure*}
	\gridline{
		\fig{fig11a.pdf}{0.4875\textwidth}{}
		\fig{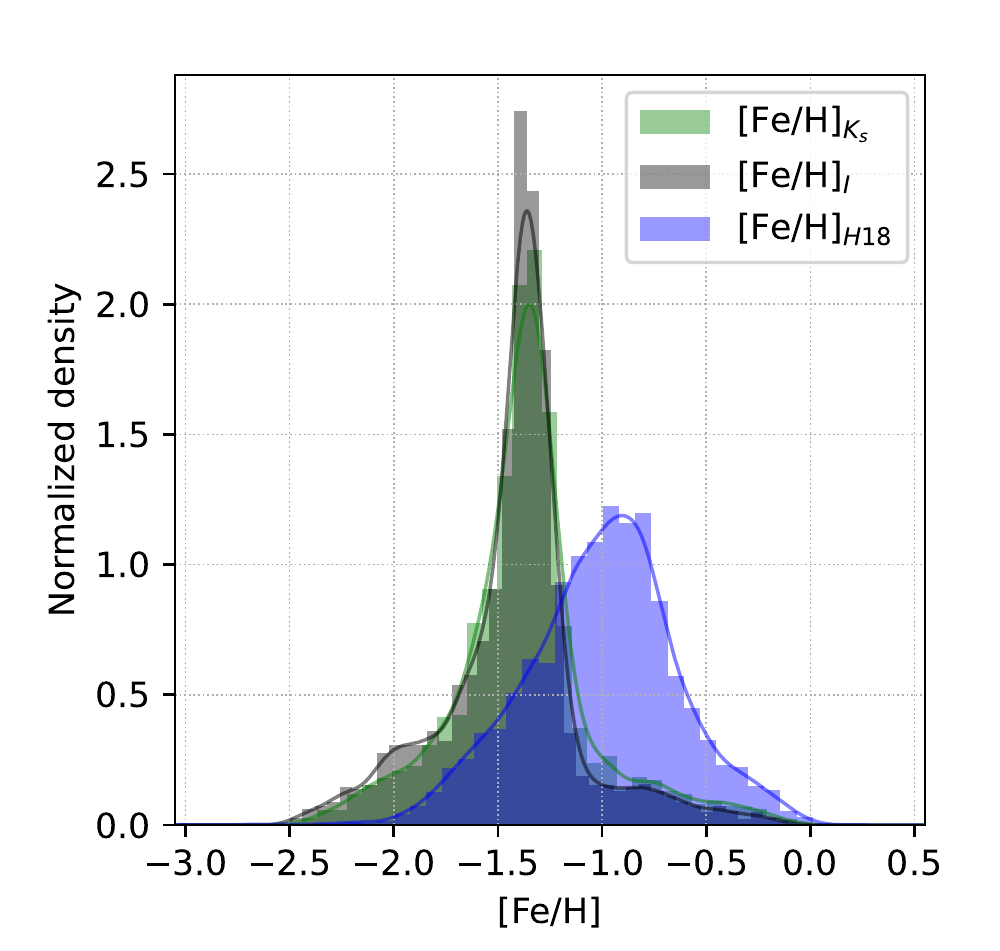}{0.39\textwidth}{}
	}
	\caption{Left: photometric metallicities predicted by our model ensemble ([Fe/H]$_{K_s}$) for the $K_s$-band target data set, plotted against metallicity estimates by the H18 method ([Fe/H]$_{\rm H18}$). The pulsation periods are color-coded, the red line denotes the identity function. Right: histograms and corresponding kernel density estimates of [Fe/H]$_{K_s}$ (green) and [Fe/H]$_{\rm H18}$ (blue) values for the target data set, and [Fe/H]$_I$ values for the development data set (gray). 
		\label{fig:kfeh_h18feh}}
\end{figure*}

\begin{figure*}
	\gridline{
		\fig{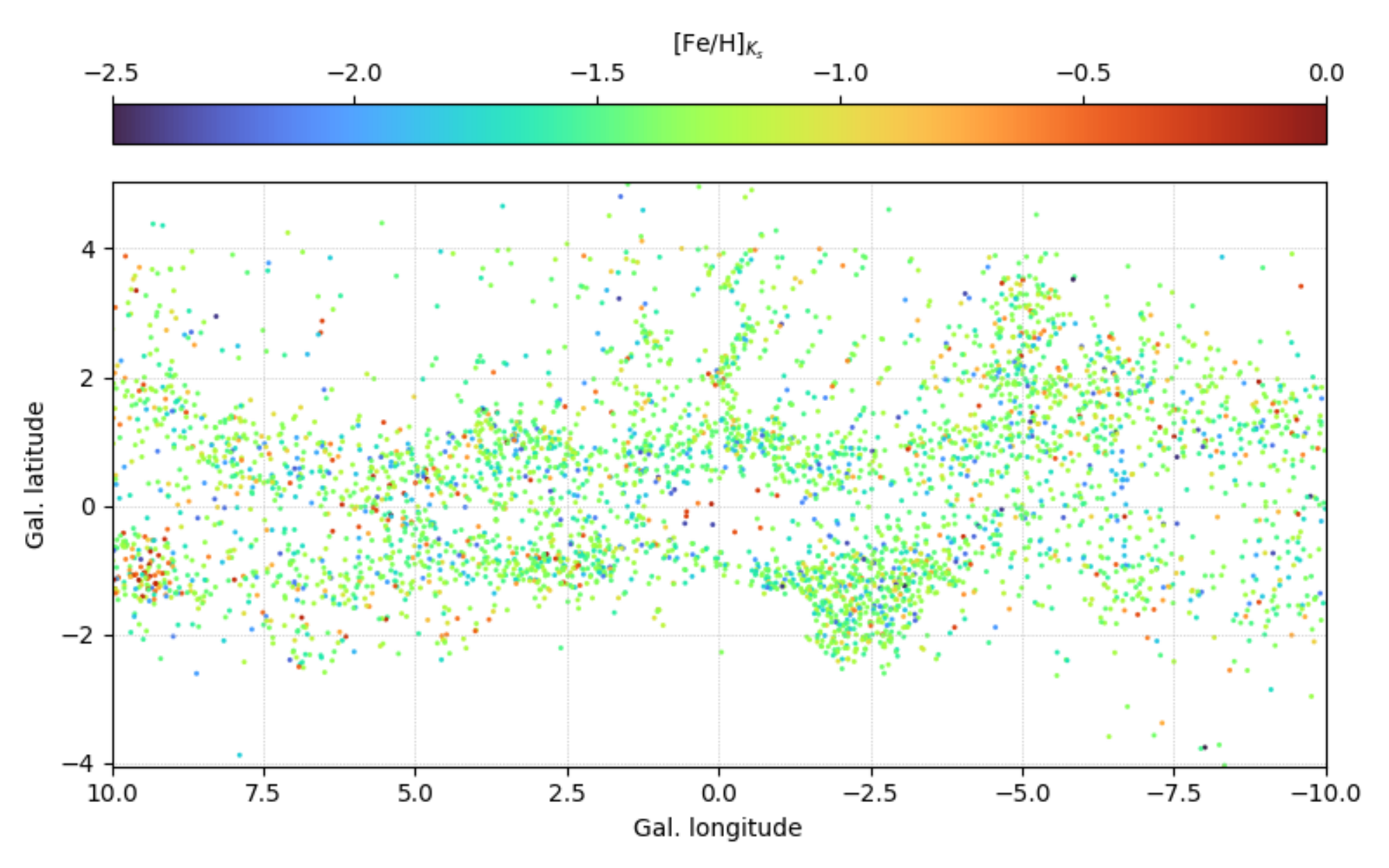}{0.8\textwidth}{}
	}
	\caption{Celestial distribution of the target data set of our Gaia $K_s$-band photometric metallicity estimator in the Galactic coordinate system. The predicted [Fe/H]$_{K_s}$ values are color-coded.
		\label{fig:kfeh_map}}
\end{figure*}

The right panel of Figure~\ref{fig:kfeh_h18feh} shows the resulting metallicity distribution of the inner bulge RRab stars, compared to the distribution of the D21 $I$-band [Fe/H] estimates of our development data set. The two distributions are very similar with their modes at $-1.35$ and $-1.36$~dex, respectively, showing a smooth continuation of the bulge MDF toward low Galactic latitudes. The figure also indicates the accuracy of both our model ensemble developed in this study, and the D20 deep-learned RRab light-curve classifier. We can observe minor differences between the two MDFs in their metal-rich and metal-poor tails. The MDF of the development set shows a very slightly enhanced metal-poor tail. Based on the OCVS only, a similar difference between the inner and outer bulge's MDF was noted by D21. The MDF of the D20 inner bulge sample also features a slightly more enhanced metal-rich tail, which can be expected, since it covers lower Galactic latitudes, where the relative fraction of metal-rich stars of the thick disk is higher. Figure~\ref{fig:kfeh_map} shows the celestial distribution of the target data set with the [Fe/H]$_{K_s}$ predictions color-coded. Around the Galactic equator, the distribution is dominated by incompleteness due to thick interstellar dust that remains impenetrable for the VISTA telescope even at two microns. The few stars detected around the Galactic Center sight-line are in fact mostly metal-rich stars in the foreground disk. We note that the linear patterns on the northern side are gaps between OGLE fields, where RRab stars that remained previously undiscovered by OGLE were detected in VVV data. The slight overdensity at $(l,b)\sim(9,-1)$ is due to an extinction window where faint metal-rich disk RRab stars behind the bulge are detected. Together with the OCVS RRab stars (for which the $I$-band [Fe/H] estimates in Table~\ref{tab:ifeh_ocvs} are preferred), this sample provides an unbiased, homogeneous sample of photometric metallicities.

Finally, we compare our results with another method for metallicity estimation from $K_s$-band light curves, that was published by \citet[][H18]{hajdu_data-driven_2018}. Following an approach similar to our study, they developed a fully connected neural network to predict the [Fe/H] of RRab stars from a 4-parameter representation of their light curves. In the latter, the phase-folded light curves were fitted with a linear combination of 4 eigenvectors obtained from a principal component analysis, and the neural network made its predictions from these 4 regression coefficients. The model was trained on $I$-band photometric metallicities obtained by the formula of \citet{smolec_metallicity_2005}. Since the latter has a positive bias with respect to our D21 formula \citep[see][]{2021ApJ...920...33D}, we can expect that the H18 model will also output systematically higher predictions compared to our LSTM neural network.  We verify this by directly comparing our [Fe/H]$_{K_s}$ predictions to those computed by the H18 model for the target data set, shown in the left panel of Fig.~\ref{fig:kfeh_h18feh}. The 
[Fe/H]$_{\rm H18}$ estimates have a positive bias of 0.4~dex on average, with a significant structure in the residual. The inner bulge MDF obtained by the H18 model is shown with blue in the right panel of Fig.~\ref{fig:kfeh_h18feh}. Similarly to the $G$-band DR2 and IB20 estimates, the metal-rich tail of the distribution is not recovered, due to its blending with the distribution's overestimated mode. The MDF is also wider due to the significantly higher scatter of the H18 predictions.

\begin{deluxetable*}{lDDDcDDDD}
	\tablecaption{$G$-band photometric metallicity estimates and basic light-curve attributes of Gaia DR2 RRab stars}
	\tablehead{
		\colhead{Gaia DR2 source\_id} & \multicolumn2c{[Fe/H]$_G$\tablenotemark{a}} & \multicolumn2c{$\sigma$([Fe/H]$_G$)\tablenotemark{b}} & \multicolumn2c{$\langle G \rangle$} & \colhead{$N_{ep.}$} & \multicolumn2c{Period [d]} & \multicolumn2c{$A_{tot.}$} & \multicolumn2c{$C_{\varphi}$} & \multicolumn2c{$S/N$}
	}
	\decimals
	\startdata
	11977018818239872  & -1.73 & 0.04 & 16.708 & 38 & 0.515610 & 0.903 & 0.881 & 125.5  \\
	11991514330833408  & -2.01 & 0.03 & 18.929 & 36 & 0.564529 & 0.824 & 0.923 & 152.0  \\
	36039689056047872  & -1.45 & 0.02 & 19.131 & 29 & 0.600486 & 0.570 & 0.880 & 108.0  \\
	36110989810386560  & -2.10 & 0.17 & 18.246 & 29 & 0.607803 & 0.775 & 0.901 & 43.2   \\
	36246122364621440  & -1.71 & 0.03 & 18.904 & 30 & 0.542566 & 0.893 & 0.886 & 177.9  \\
	\enddata
	\tablenotetext{a}{Mean of predictions from a model ensemble.}
	\tablenotetext{b}{Standard deviation of predictions from a model ensemble.}
	\tablecomments{This table is available in its entirety in machine-readable form.}
\end{deluxetable*}\label{tab:gfeh}

\begin{deluxetable*}{lhccDDDcDDDD}
	\tablecaption{$K_s$-band photometric metallicity estimates and basic light-curve attributes of the VVV bulge RRab stars discovered by \citet{dekany_near-infrared_2020}}
	\tablehead{
		\colhead{ID\tablenotemark{a}} & \nocolhead{VVVID} & \colhead{R.A.\tablenotemark{b} [hms]} & \colhead{DEC.\tablenotemark{b} [dms]} & \multicolumn2c{[Fe/H]$_{K_s}$\tablenotemark{c}} & \multicolumn2c{$\sigma$([Fe/H]$_{K_s}$)\tablenotemark{d}} & \multicolumn2c{$\langle K_s \rangle$} & \colhead{$N_{ep.}$} & \multicolumn2c{Period [d]} & \multicolumn2c{$A_{tot.}$} & \multicolumn2c{$C_{\varphi}$} & \multicolumn2c{$S/N$}
	}
	\decimals
	\startdata
	7    &  3690341459 &  17:05:27.27 &  -34:33:50.9 &  -1.27 &  0.07 &  14.657 &    117  &  0.643188 &  0.171  &    0.939 &  78.5 \\
	27   &  3690245844 &  17:06:58.67 &  -34:59:07.7 &  -0.40 &  0.05 &  15.538 &     64  &  0.440602 &  0.245  &    0.938 &  56.8 \\
	36   &  3690652405 &  17:07:41.14 &  -34:18:33.7 &  -1.48 &  0.05 &  14.421 &     64  &  0.769114 &  0.157  &    0.933 &  82.2 \\
	64   &  3700148111 &  17:08:58.02 &  -33:40:19.8 &  -1.81 &  0.12 &  14.808 &     65  &  0.737811 &  0.143  &    0.915 &  46.3 \\
	108  &  3550383138 &  17:10:26.66 &  -35:14:50.0 &  -1.30 &  0.02 &  14.573 &     65  &  0.632416 &  0.181  &    0.924 &  91.8 \\
	\enddata
	\tablenotetext{a}{Identifier from \citet{dekany_near-infrared_2020}.}
	\tablenotetext{b}{Coordinates of J2000.0 epoch.}
	\tablenotetext{c}{Mean of predictions from a model ensemble.}
	\tablenotetext{d}{Standard deviation of predictions from a model ensemble.}
	\tablecomments{This table is available in its entirety in machine-readable form.}
\end{deluxetable*}\label{tab:kfeh}

\section{Summary and Conclusions}

In this study, we have provided a new method for the estimation  of the metallicity of RRab stars from their Gaia $G$-band and near-infrared $K_s$-band light curves. In our approach, we have leveraged deep learning to accurately transfer a carefully calibrated photometric [Fe/H] prediction formula to additional wavebands at the cost of minimal additional noise. To achieve this, we used the $I$-band D21 formula, which had been calibrated to HDS metallicities, to compute photometric metallicities for a large number of objects that also have light curves in the $G$ and/or $K_s$ bands. Subsequently, these metallicities served as response variables in separate regression problems of the $G$ and $K_s$ light curves, solved by state-of-the-art recurrent neural networks. The resulting model ensembles have very high predictive performance measured by cross-validation, with a (weighted) mean absolute error of only 0.1~dex in predicting the $I$-band photometric metallicities. They are also able to recover the metallicity distributions of the validation data sets without bias, demonstrating that our models generalize well to new data.

Comparing our results with predictions of various other methods from the literature, we found that earlier methods generally overestimate the metallicity with respect to the D21 formula, and consequently the LSTM models of this study, which were trained on [Fe/H]$_I$ values from the former. Systematic errors in earlier photometric [Fe/H] formulae resulted in positive biases of $\gtrsim0.4$~\,dex and $\gtrsim0.25$~dex in the modes of the bulge's and the halo's MDFs, respectively. We found that the [Fe/H] values provided with the Gaia DR2 RR~Lyrae catalog are particularly discrepant. We suspect that these disagreements stem from multiple sources, including biases in regressions due to small and imbalanced calibration data sets, in the transformations of Fourier parameters between wavebands, and in the low-resolution spectroscopic determinations of the metallicity used for the calibration of most of the earlier methods.

We deployed our predictive models to compute the [Fe/H] for numerous Galactic RRab stars observed by the OGLE, Gaia, and VVV surveys. The resulting catalogs of photometric metallicities are provided in the electronic versions of Tables~\ref{tab:ifeh_ocvs}, \ref{tab:gfeh}, and \ref{tab:kfeh}, and can be directly incorporated in various future studies of Galactic archeology. We also make our software code \citep{rrl_feh_nn} publicly available online\footnote{\url{https://zenodo.org/record/6576131}}, in order to facilitate the further deployment of the neural networks, as well as to allow their eventual re-training and further optimization on other data sets.

The approach of metallicity estimation presented in this paper should be easily applicable in the future to other surveys as well, provided that large enough data sets with well-sampled light curves in multiple photometric wavebands are available for training the neural networks. We also anticipate that the upcoming Gaia data releases will have improved light-curve sampling and accuracy, and they will include many more RR~Lyrae stars, allowing the future improvement of our models' predictive performance. Upon the availability of  these data releases, the timely update of our method will be straightforward by retraining our published neural networks on the new data sets. 

Our photometric prediction method makes it straightforward to obtain consistent metallicities for very large combined samples of RRab stars, spanning across the footprints of various sky surveys conducted in different wavebands. The resulting homogeneous metallicity estimates, combined with astrometric or photometric distances, make RRab stars outstanding chemo-dynamical tracer objects in the era of data-driven astronomy.

\begin{acknowledgments}
The authors were supported by the Deutsche Forschungsgemeinschaft (DFG, German Research Foundation) -- Project-ID 138713538 -- SFB 881 (``The Milky Way System'', subproject A03).
\end{acknowledgments}

\vspace{5mm}

\software{
	numpy \citep{harris2020array},
	scipy \citep{2020SciPy-NMeth},
	scikit-learn \citep{scikit-learn},
	TensorFlow \citep{tensorflow2015-whitepaper},
	lcfit \citep{lcfit},
	rrl\textunderscore feh\textunderscore nn \citep{rrl_feh_nn}
}

\bibliography{references}{}

\begin{thebibliography}{}
\expandafter\ifx\csname natexlab\endcsname\relax\def\natexlab#1{#1}\fi
\providecommand{\url}[1]{\href{#1}{#1}}
\providecommand{\dodoi}[1]{doi:~\href{http://doi.org/#1}{\nolinkurl{#1}}}
\providecommand{\doeprint}[1]{\href{http://ascl.net/#1}{\nolinkurl{http://ascl.net/#1}}}
\providecommand{\doarXiv}[1]{\href{https://arxiv.org/abs/#1}{\nolinkurl{https://arxiv.org/abs/#1}}}

\bibitem[{Abadi {et~al.}(2015)Abadi, Agarwal, Barham, Brevdo, Chen, Citro,
  Corrado, Davis, Dean, Devin, Ghemawat, Goodfellow, Harp, Irving, Isard, Jia,
  Jozefowicz, Kaiser, Kudlur, Levenberg, Man\'{e}, Monga, Moore, Murray, Olah,
  Schuster, Shlens, Steiner, Sutskever, Talwar, Tucker, Vanhoucke, Vasudevan,
  Vi\'{e}gas, Vinyals, Warden, Wattenberg, Wicke, Yu, \&
  Zheng}]{tensorflow2015-whitepaper}
Abadi, M., Agarwal, A., Barham, P., {et~al.} 2015, {TensorFlow}: Large-Scale
  Machine Learning on Heterogeneous Systems.
\newblock \url{https://www.tensorflow.org/}

\bibitem[{{An} {et~al.}(2013){An}, {Beers}, {Johnson}, {Pinsonneault}, {Lee},
  {Bovy}, {Ivezi{\'c}}, {Carollo}, \& {Newby}}]{2013ApJ...763...65A}
{An}, D., {Beers}, T.~C., {Johnson}, J.~A., {et~al.} 2013, \apj, 763, 65,
  \dodoi{10.1088/0004-637X/763/1/65}

\bibitem[{Andrievsky {et~al.}(2018)Andrievsky, Wallerstein, Korotin, Lyashko,
  Kovtyukh, Tsymbal, Davis, Gomez, Huang, \&
  Farrell}]{andrievsky_relationship_2018}
Andrievsky, S., Wallerstein, G., Korotin, S., {et~al.} 2018, Publications of
  the Astronomical Society of the Pacific, 130, 024201,
  \dodoi{10.1088/1538-3873/aa9783}

\bibitem[{{Bhardwaj}(2022)}]{2022arXiv220206982B}
{Bhardwaj}, A. 2022, arXiv e-prints, arXiv:2202.06982.
\newblock \doarXiv{2202.06982}

\bibitem[{Chadid {et~al.}(2017)Chadid, Sneden, \&
  Preston}]{chadid_spectroscopic_2017}
Chadid, M., Sneden, C., \& Preston, G.~W. 2017, The Astrophysical Journal, 835,
  187, \dodoi{10.3847/1538-4357/835/2/187}

\bibitem[{Clementini {et~al.}(1995)Clementini, Carretta, Gratton, Merighi,
  Mould, \& McCarthy}]{clementini_composition_1995}
Clementini, G., Carretta, E., Gratton, R., {et~al.} 1995, The Astronomical
  Journal, 110, 2319, \dodoi{10.1086/117692}

\bibitem[{Clementini {et~al.}(2019)Clementini, Ripepi, Molinaro, Garofalo,
  Muraveva, Rimoldini, Guy, Jevardat~de Fombelle, Nienartowicz, Marchal,
  Audard, Holl, Leccia, Marconi, Musella, Mowlavi, Lecoeur-Taibi, Eyer,
  De~Ridder, Regibo, Sarro, Szabados, Evans, \& Riello}]{clementini_gaia_2019}
Clementini, G., Ripepi, V., Molinaro, R., {et~al.} 2019, Astronomy \&
  Astrophysics, 622, A60, \dodoi{10.1051/0004-6361/201833374}

\bibitem[{{Crestani} {et~al.}(2021){Crestani}, {Fabrizio}, {Braga}, {Sneden},
  {Preston}, {Ferraro}, {Iannicola}, {Bono}, {Alves-Brito}, {Nonino},
  {D'Orazi}, {Inno}, {Monelli}, {Storm}, {Altavilla}, {Chaboyer}, {Dall'Ora},
  {Fiorentino}, {Gilligan}, {Grebel}, {Lala}, {Lemasle}, {Marengo}, {Marinoni},
  {Marrese}, {Mart{\'\i}nez-V{\'a}zquez}, {Matsunaga}, {Mullen}, {Neeley},
  {Prudil}, {da Silva}, {Stetson}, {Th{\'e}venin}, {Valenti}, {Walker}, \&
  {Zoccali}}]{crestani_deltaS}
{Crestani}, J., {Fabrizio}, M., {Braga}, V.~F., {et~al.} 2021, \apj, 908, 20,
  \dodoi{10.3847/1538-4357/abd183}

\bibitem[{{D{\'e}k{\'a}ny}(2022{\natexlab{a}})}]{lcfit}
{D{\'e}k{\'a}ny}, I. 2022{\natexlab{a}}, lcfit: A python package for the
  regression of periodic time series, 1.0.0,  Zenodo,
  \dodoi{10.5281/ZENODO.6576222}

\bibitem[{{D{\'e}k{\'a}ny}(2022{\natexlab{b}})}]{rrl_feh_nn}
---. 2022{\natexlab{b}}, {rrl\textunderscore feh\textunderscore nn:
  Deep-learned metallicity estimation of RR Lyrae-type pulsating stars from
  their $G$ and $K_s$ light curves}, 1.0.0,  Zenodo,
  \dodoi{10.5281/ZENODO.6576131}

\bibitem[{{D{\'e}k{\'a}ny} {et~al.}(2021){D{\'e}k{\'a}ny}, {Grebel}, \&
  {Pojma{\'n}ski}}]{2021ApJ...920...33D}
{D{\'e}k{\'a}ny}, I., {Grebel}, E.~K., \& {Pojma{\'n}ski}, G. 2021, \apj, 920,
  33, \dodoi{10.3847/1538-4357/ac106f}

\bibitem[{Dékány \& Grebel(2020)}]{dekany_near-infrared_2020}
Dékány, I., \& Grebel, E.~K. 2020, The Astrophysical Journal, 898, 46,
  \dodoi{10.3847/1538-4357/ab9d87}

\bibitem[{Dékány {et~al.}(2018)Dékány, Hajdu, Grebel, Catelan, Elorrieta,
  Eyheramendy, Majaess, \& Jordán}]{dekany_near-infrared_2018}
Dékány, I., Hajdu, G., Grebel, E.~K., {et~al.} 2018, The Astrophysical
  Journal, 857, 54, \dodoi{10.3847/1538-4357/aab4fa}

\bibitem[{{Emerson} {et~al.}(2004){Emerson}, {Irwin}, {Lewis}, {Hodgkin},
  {Evans}, {Bunclark}, {McMahon}, {Hambly}, {Mann}, {Bond}, {Sutorius}, {Read},
  {Williams}, {Lawrence}, \& {Stewart}}]{2004SPIE.5493..401E}
{Emerson}, J.~P., {Irwin}, M.~J., {Lewis}, J., {et~al.} 2004, in Society of
  Photo-Optical Instrumentation Engineers (SPIE) Conference Series, Vol. 5493,
  Optimizing Scientific Return for Astronomy through Information Technologies,
  ed. P.~J. {Quinn} \& A.~{Bridger}, 401--410, \dodoi{10.1117/12.551582}

\bibitem[{Fernley \& Barnes(1996)}]{fernley_metal_1996}
Fernley, J., \& Barnes, T.~G. 1996, Astronomy \& Astrophysics, 312, 957.
\newblock \url{http://adsabs.harvard.edu/abs/1996A%26A...312..957F}

\bibitem[{For {et~al.}(2011)For, Sneden, \& Preston}]{for_chemical_2011}
For, B.-Q., Sneden, C., \& Preston, G.~W. 2011, The Astrophysical Journal
  Supplement Series, 197, 29, \dodoi{10.1088/0067-0049/197/2/29}

\bibitem[{Govea {et~al.}(2014)Govea, Gomez, Preston, \&
  Sneden}]{govea_chemical_2014}
Govea, J., Gomez, T., Preston, G.~W., \& Sneden, C. 2014, The Astrophysical
  Journal, 782, 59, \dodoi{10.1088/0004-637X/782/2/59}

\bibitem[{{Hajdu} {et~al.}(2020){Hajdu}, {D{\'e}k{\'a}ny}, {Catelan}, \&
  {Grebel}}]{2020ExA....49..217H}
{Hajdu}, G., {D{\'e}k{\'a}ny}, I., {Catelan}, M., \& {Grebel}, E.~K. 2020,
  Experimental Astronomy, 49, 217, \dodoi{10.1007/s10686-020-09661-0}

\bibitem[{Hajdu {et~al.}(2018)Hajdu, Dékány, Catelan, Grebel, \&
  Jurcsik}]{hajdu_data-driven_2018}
Hajdu, G., Dékány, I., Catelan, M., Grebel, E.~K., \& Jurcsik, J. 2018, The
  Astrophysical Journal, 857, 55, \dodoi{10.3847/1538-4357/aab4fd}

\bibitem[{Harris {et~al.}(2020)Harris, Millman, van~der Walt, Gommers,
  Virtanen, Cournapeau, Wieser, Taylor, Berg, Smith, Kern, Picus, Hoyer, van
  Kerkwijk, Brett, Haldane, del R{'{\i}}o, Wiebe, Peterson,
  G{'{e}}rard-Marchant, Sheppard, Reddy, Weckesser, Abbasi, Gohlke, \&
  Oliphant}]{harris2020array}
Harris, C.~R., Millman, K.~J., van~der Walt, S.~J., {et~al.} 2020, Nature, 585,
  357, \dodoi{10.1038/s41586-020-2649-2}

\bibitem[{Hochreiter \& Schmidhuber(1997)}]{lstm_hochreiter_schmidhuber}
Hochreiter, S., \& Schmidhuber, J. 1997, Neural Computation, 9, 1735,
  \dodoi{10.1162/neco.1997.9.8.1735}

\bibitem[{Houdt {et~al.}(2020)Houdt, Mosquera, \& N{\'a}poles}]{Houdt2020ARO}
Houdt, G.~V., Mosquera, C., \& N{\'a}poles, G. 2020, Artificial Intelligence
  Review, 1

\bibitem[{Iorio \& Belokurov(2020)}]{iorio_chemo-kinematics_2020}
Iorio, G., \& Belokurov, V. 2020, arXiv e-prints, 2008, arXiv:2008.02280.
\newblock \url{http://adsabs.harvard.edu/abs/2020arXiv200802280I}

\bibitem[{{Jurcsik} {et~al.}(2018){Jurcsik}, {Hajdu}, {D{\'e}k{\'a}ny},
  {Nuspl}, {Catelan}, \& {Grebel}}]{2018MNRAS.475.4208J}
{Jurcsik}, J., {Hajdu}, G., {D{\'e}k{\'a}ny}, I., {et~al.} 2018, \mnras, 475,
  4208, \dodoi{10.1093/mnras/sty112}

\bibitem[{Jurcsik \& Kov\'acs(1996)}]{jurcsik_determination_1996}
Jurcsik, J., \& Kov\'acs, G. 1996, Astronomy \& Astrophysics, 312, 111

\bibitem[{{Kingma} \& {Ba}(2014)}]{2014arXiv1412.6980K}
{Kingma}, D.~P., \& {Ba}, J. 2014, arXiv e-prints, arXiv:1412.6980.
\newblock \doarXiv{1412.6980}

\bibitem[{Lambert {et~al.}(1996)Lambert, Heath, Lemke, \&
  Drake}]{lambert_chemical_1996}
Lambert, D.~L., Heath, J.~E., Lemke, M., \& Drake, J. 1996, The Astrophysical
  Journal Supplement Series, 103, 183, \dodoi{10.1086/192274}

\bibitem[{Layden(1994)}]{layden_metallicities_1994}
Layden, A.~C. 1994, The Astronomical Journal, 108, 1016, \dodoi{10.1086/117132}

\bibitem[{Liu {et~al.}(2013)Liu, Zhao, Chen, Takeda, \&
  Honda}]{liu_abundances_2013}
Liu, S., Zhao, G., Chen, Y.-Q., Takeda, Y., \& Honda, S. 2013, Research in
  Astronomy and Astrophysics, 13, 1307, \dodoi{10.1088/1674-4527/13/11/003}

\bibitem[{Minniti {et~al.}(2010)Minniti, Lucas, Emerson, Saito, Hempel,
  Pietrukowicz, Ahumada, Alonso, Alonso-Garcia, Arias, Bandyopadhyay, Barbá,
  Barbuy, Bedin, Bica, Borissova, Bronfman, Carraro, Catelan, Clariá, Cross,
  de~Grijs, Dékány, Drew, Fari{\textbackslash}textasciitilde~na, Feinstein,
  Fernández~Lajús, Gamen, Geisler, Gieren, Goldman, Gonzalez, Gunthardt,
  Gurovich, Hambly, Irwin, Ivanov, Jordán, Kerins, Kinemuchi, Kurtev,
  López-Corredoira, Maccarone, Masetti, Merlo, Messineo, Mirabel, Monaco,
  Morelli, Padilla, Palma, Parisi, Pignata, Rejkuba, Roman-Lopes, Sale,
  Schreiber, Schröder, Smith, Sodré, Soto, Tamura, Tappert, Thompson, Toledo,
  Zoccali, \& Pietrzynski}]{minniti_vista_2010}
Minniti, D., Lucas, P.~W., Emerson, J.~P., {et~al.} 2010, {\textbackslash}na,
  15, 433, \dodoi{10.1016/j.newast.2009.12.002}

\bibitem[{{Mullen} {et~al.}(2021){Mullen}, {Marengo},
  {Mart{\'\i}nez-V{\'a}zquez}, {Neeley}, {Bono}, {Dall'Ora}, {Chaboyer},
  {Th{\'e}venin}, {Braga}, {Crestani}, {Fabrizio}, {Fiorentino}, {Gilligan},
  {Monelli}, \& {Stetson}}]{mullen_metallicity_2021}
{Mullen}, J.~P., {Marengo}, M., {Mart{\'\i}nez-V{\'a}zquez}, C.~E., {et~al.}
  2021, \apj, 912, 144, \dodoi{10.3847/1538-4357/abefd4}

\bibitem[{Nemec {et~al.}(2013)Nemec, Cohen, Ripepi, Derekas, Moskalik, Sesar,
  Chadid, \& Bruntt}]{nemec_metal_2013}
Nemec, J.~M., Cohen, J.~G., Ripepi, V., {et~al.} 2013, The Astrophysical
  Journal, 773, 181, \dodoi{10.1088/0004-637X/773/2/181}

\bibitem[{Ngeow {et~al.}(2016)Ngeow, Yu, Bellm, Yang, Chang, Miller, Laher,
  Surace, \& Ip}]{ngeow_palomar_2016}
Ngeow, C.-C., Yu, P.-C., Bellm, E., {et~al.} 2016, The Astrophysical Journal
  Supplement Series, 227, 30, \dodoi{10.3847/1538-4365/227/2/30}

\bibitem[{Pancino {et~al.}(2015)Pancino, Britavskiy, Romano, Cacciari,
  Mucciarelli, \& Clementini}]{pancino_chemical_2015}
Pancino, E., Britavskiy, N., Romano, D., {et~al.} 2015, Monthly Notices of the
  Royal Astronomical Society, 447, 2404, \dodoi{10.1093/mnras/stu2616}

\bibitem[{Pedregosa {et~al.}(2011)Pedregosa, Varoquaux, Gramfort, Michel,
  Thirion, Grisel, Blondel, Prettenhofer, Weiss, Dubourg, Vanderplas, Passos,
  Cournapeau, Brucher, Perrot, \& Duchesnay}]{scikit-learn}
Pedregosa, F., Varoquaux, G., Gramfort, A., {et~al.} 2011, Journal of Machine
  Learning Research, 12, 2825

\bibitem[{{Prudil} \& {Skarka}(2017)}]{2017MNRAS.466.2602P}
{Prudil}, Z., \& {Skarka}, M. 2017, \mnras, 466, 2602,
  \dodoi{10.1093/mnras/stw3231}

\bibitem[{Savino {et~al.}(2020)Savino, Koch, Prudil, Kunder, \&
  Smolec}]{savino_age_2020}
Savino, A., Koch, A., Prudil, Z., Kunder, A., \& Smolec, R. 2020, Astronomy \&
  Astrophysics, 641, A96, \dodoi{10.1051/0004-6361/202038305}

\bibitem[{Skowron {et~al.}(2016)Skowron, Soszyński, Udalski, Szymański,
  Pietrukowicz, Skowron, Poleski, Wyrzykowski, Ulaczyk, Kozłowski, Mróz, \&
  Pawlak}]{skowron_ogle-ing_2016}
Skowron, D.~M., Soszyński, I., Udalski, A., {et~al.} 2016, Acta Astronomica,
  66, 269.
\newblock \url{http://adsabs.harvard.edu/abs/2016AcA....66..269S}

\bibitem[{Smolec(2005)}]{smolec_metallicity_2005}
Smolec, R. 2005, Acta Astronomica, 55, 59

\bibitem[{{Smolec}(2016)}]{2016pas..conf...22S}
{Smolec}, R. 2016, in 37th Meeting of the Polish Astronomical Society, ed.
  A.~{R{\'o}{\.z}a{\'n}ska} \& M.~{Bejger}, Vol.~3, 22--25.
\newblock \doarXiv{1603.01252}

\bibitem[{Sneden {et~al.}(2017)Sneden, Preston, Chadid, \&
  Adamów}]{sneden_rrc_2017}
Sneden, C., Preston, G.~W., Chadid, M., \& Adamów, M. 2017,
  {\textbackslash}apj, 848, 68, \dodoi{10.3847/1538-4357/aa8b10}

\bibitem[{Soszyński {et~al.}(2019)Soszyński, Udalski, Wrona, Szymański,
  Pietrukowicz, Skowron, Skowron, Poleski, Kozłowski, Mróz, Ulaczyk, Rybicki,
  Iwanek, \& Gromadzki}]{soszynski_over_2019}
Soszyński, I., Udalski, A., Wrona, M., {et~al.} 2019, Acta Astronomica, 69,
  321, \dodoi{10.32023/0001-5237/69.4.2}

\bibitem[{Srivastava {et~al.}(2014)Srivastava, Hinton, Krizhevsky, Sutskever,
  \& Salakhutdinov}]{JMLR:v15:srivastava14a}
Srivastava, N., Hinton, G., Krizhevsky, A., Sutskever, I., \& Salakhutdinov, R.
  2014, Journal of Machine Learning Research, 15, 1929.
\newblock \url{http://jmlr.org/papers/v15/srivastava14a.html}

\bibitem[{{Tanakul} \& {Sarajedini}(2018)}]{2018MNRAS.478.4590T}
{Tanakul}, N., \& {Sarajedini}, A. 2018, \mnras, 478, 4590,
  \dodoi{10.1093/mnras/sty1311}

\bibitem[{Udalski {et~al.}(2015)Udalski, Szymański, \&
  Szymański}]{udalski_ogle-iv_2015}
Udalski, A., Szymański, M.~K., \& Szymański, G. 2015, Acta Astronomica, 65,
  1.
\newblock \url{http://adsabs.harvard.edu/abs/2015AcA....65....1U}

\bibitem[{{Vida} {et~al.}(2021){Vida}, {B{\'o}di}, {Szklen{\'a}r}, \&
  {Seli}}]{2021A&A...652A.107V}
{Vida}, K., {B{\'o}di}, A., {Szklen{\'a}r}, T., \& {Seli}, B. 2021, \aap, 652,
  A107, \dodoi{10.1051/0004-6361/202141068}

\bibitem[{Virtanen {et~al.}(2020)Virtanen, Gommers, Oliphant, Haberland, Reddy,
  Cournapeau, Burovski, Peterson, Weckesser, Bright, {van der Walt}, Brett,
  Wilson, Millman, Mayorov, Nelson, Jones, Kern, Larson, Carey, Polat, Feng,
  Moore, {VanderPlas}, Laxalde, Perktold, Cimrman, Henriksen, Quintero, Harris,
  Archibald, Ribeiro, Pedregosa, {van Mulbregt}, \& {SciPy 1.0
  Contributors}}]{2020SciPy-NMeth}
Virtanen, P., Gommers, R., Oliphant, T.~E., {et~al.} 2020, Nature Methods, 17,
  261, \dodoi{10.1038/s41592-019-0686-2}

\bibitem[{{Youakim} {et~al.}(2020){Youakim}, {Starkenburg}, {Martin},
  {Matijevi{\v{c}}}, {Aguado}, {Allende Prieto}, {Arentsen}, {Bonifacio},
  {Carlberg}, {Gonz{\'a}lez Hern{\'a}ndez}, {Hill}, {Kordopatis}, {Lardo},
  {Navarro}, {Jablonka}, {S{\'a}nchez Janssen}, {Sestito}, {Thomas}, \&
  {Venn}}]{2020MNRAS.492.4986Y}
{Youakim}, K., {Starkenburg}, E., {Martin}, N.~F., {et~al.} 2020, \mnras, 492,
  4986, \dodoi{10.1093/mnras/stz3619}

\bibitem[{Yu {et~al.}(2019)Yu, Si, Hu, \& Zhang}]{yu_rnn_review}
Yu, Y., Si, X., Hu, C., \& Zhang, J. 2019, Neural Computation, 31, 1235,
  \dodoi{10.1162/neco_a_01199}

\end{thebibliography}
\bibliographystyle{aasjournal}



\end{document}